\documentclass[journal]{IEEEtran}
\usepackage{cite}
\usepackage{url}
\usepackage{color,soul}

\usepackage{tikz,forest}
\ifCLASSINFOpdf
\usepackage{graphicx}
\usepackage{tabularx}
\usepackage{amssymb}
\usepackage{float}
\graphicspath{{./pdf/}{./jpeg/}}
\else
\fi

\usepackage{algorithm}
\usepackage{algorithmicx}
\usepackage[noend]{algpseudocode}
\let\oldemptyset\emptyset
\let\emptyset\varnothing

\makeatletter
\renewcommand{\ALG@beginalgorithmic}{\small}

\def\BState{\State\hskip-\ALG@thistlm}
\makeatother

\usepackage{array}
\ifCLASSOPTIONcompsoc
  \usepackage[caption=false,font=normalsize,labelfont=sf,textfont=sf]{subfig}
\else
  \usepackage[caption=false,font=footnotesize]{subfig}
\fi
\usepackage[cmex10]{amsmath}
\usetikzlibrary{calc}
\usetikzlibrary{positioning}
\usetikzlibrary{decorations.text}
\usetikzlibrary{arrows.meta}
\usepackage{multirow}

\begin{document}
\title{Impact of Adaptive Consistency on Distributed SDN Applications: An Empirical Study}

\author{Ermin~Sakic,~\IEEEmembership{Student Member,~IEEE,}
        and Wolfgang~Kellerer,~\IEEEmembership{Senior Member,~IEEE}%
	        %and~Jane~Doe,~\IEEEmembership{Life~Fellow,~IEEE}% <-this % stops a space
		\thanks{E. Sakic is with the Department
		of Electrical and Computer Engineering, University of Technology Munich, Germany;
		and Siemens AG, Munich, Germany, E-Mail: (ermin.sakic@\{tum.de, siemens.com\}).}%
		\thanks{W. Kellerer is with the Department of Electrical and Computer Engineering, 
		University of Technology Munich, Germany, E-Mail: (wolfgang.kellerer@tum.de).}% <-this % stops a space
		%\thanks{Manuscript received May XX, 2017.}}

		\vspace{-7mm} 
		}
%\author{\IEEEauthorblockN{Ermin Sakic* and Wolfgang Kellerer**}
%	\IEEEauthorblockA{* \{Technical University of Munich, Siemens AG\}, Munich, Germany, ermin.sakic@\{tum.de, siemens.com\}\\
%	** Technical University of Munich, Munich, Germany, wolfgang.kellerer@tum.de\\
%}
%}
\newcommand{\todo}[1]{\textcolor{red}{#1}\PackageWarning{TODO:}{#1!}}

%\author{\IEEEauthorblockN{Ermin Sakic}
%\IEEEauthorblockA{Technical University of Munich, Siemens AG\\
%Munich, Germany\\
%Email: ermin.sakic@\{tum.de, siemens.com\}}
%\and
%\IEEEauthorblockN{Wolfgang Kellerer}
%\IEEEauthorblockA{Technical University of Munich\\
%Munich, Germany\\
%Email: wolfgang.kellerer@tum.de}
%}


%Telephone: (800) 555--1212, Fax: (888) 555--1212}


\maketitle

\begin{abstract}
	Scalability of the control plane in a Software Defined Network (SDN) is enabled by means of decentralization of the decision-making logic, i.e. by replication of controller functions to physically or virtually dislocated controller replicas. Replication of a centralized controller state also enables the protection against controller failures by means of primary and backup replicas responsible for managing the underlying SDN data plane devices. In this work, we investigate the effect of the the deployed consistency model on \emph{scalability} and \emph{correctness} metrics of the SDN control plane. In particular, we compare the \emph{strong} and \emph{eventual} consistency, and make a case for a novel \emph{adaptive} consistency approach. The existing controller platforms rely on either strong or eventual consistency mechanisms in their state distribution. We show how an adaptive consistency model offers the scalability benefits in terms of the total request-handling throughput and response time, in contrast to the strong consistency model. We also outline how the adaptive consistency approach can provide for correctness semantics, that are unachievable with the eventual consistency paradigm in practice. The adaptability of our approach provides a balanced and tunable trade-off of scalability and correctness for the SDN application implemented on top of the adaptive framework. To validate our assumptions, we evaluate and compare the different approaches in an emulated testbed with an example of a load balancer controller application. The experimental setup comprises up to five extended OpenDaylight controller instances and two network topologies from the area of service provider and data center networks.
\end{abstract}


% \fi
\IEEEpeerreviewmaketitle

\section{Introduction}
The SDN paradigm aims at centralizing the network logic in a decision-making entity known as the \emph{SDN controller}. The concept of knowledge centralization has recently gained traction for its potential advantages in the abstraction and added simplicity of network control and management operations \cite{sardis2016can}. %The standardized network control interactions between the SDN controller and forwarding devices allow for the encapsulation of continuous tasks related to complex network state knowledge collection. The northbound interfaces of the SDN controller enable a simple but powerful access to the controller APIs and the retrieval of the internal controller's knowledge for consumption in external applications. 
The centralization of the controller's knowledge state, however, introduces two new challenges: the \emph{single-point-of-failure (SPOF)} and the \emph{scalability} of the control plane \cite{sdn_survey}. A number of approaches have been proposed in literature to alleviate the SPOF \cite{hyperflow, koponen2010onix, onos, katta2015ravana} issue, with the major approaches relying on direct state- and function-replication across the \emph{replicas} of the SDN controller cluster. 

%With the \emph{strong consistency model} (SC), 

With the concept of state replication, the SDN controller instances replicate their data store contents to other members that take part in a logical \emph{controller cluster}, using a state distribution protocol of choice (e.g. RAFT \cite{raft, sakicresponse}). When a failure of a controller is suspected, another replica from its \emph{cluster} is able to take over and continue to serve future application's requests. The selection of the \emph{consistency model} leveraged by the replication process affects the incurred synchronization overhead in terms of the resulting packet load, the experienced commit response times and the processing order of commits.

In the \emph{Strong Consistency} model (SC), each consecutive operation that modifies the internal state of the controller is serialized and confirmed by a quorum of replicas, before forwarding the state and processing subsequent transactions. In the leader-based SC approaches (e.g. in RAFT), all requests are serialized by a cluster \emph{leader}, in order to provide for a consistent data store view across all cluster \emph{followers}. Thus, with SC, a large distributed system consisting of multiple controller replicas, is effectively constrained into a monolithic system where each data store modification incurs a minimum of two message rounds and a linear message complexity (in the case of a stable leader) in order to synchronize the controller views \cite{sakicresponse, suh2017toward, suh2016performance}.

In the \emph{Eventual Consistency} (EC) model \cite{levin2012logically, panda2017scl, hyperflow}, state transitions may be delayed or reordered for an arbitrary period of time. In EC, message updates are advertised in a single round and with linear message complexity. From SDN controller perspective, each controller instance in EC is able to autonomously service the client requests. The updates to the internal data store are thus non-blocking and are executed without incurring an additional delay in SDN application's processing time \cite{suh2017toward}. However, in the EC the missing constraint of state serialization potentially leads to write conflicts and inefficient decision-making \cite{levin2012logically}. 

Recent works have introduced the paradigm of \emph{Adaptive Consistency} (AC) \cite{aslan2016adaptive, sakicadaptive}. In general, AC realizes the state synchronization as a non-blocking task. However, after exceeding a configurable number of maximum concurrent per-replica state-modifications, an AC system blocks further updates until all replicas have synchronized to a common state \cite{sakicadaptive}. If the system detects that the \emph{staleness} constraints of an SDN application may be violated by a concurrent state-modification, it blocks the future state modifications until the state consistency across all replicas is reestablished. Additionally, AC autonomously adapts the consistency level metric of the system. This adaptation advocates an asynchronous state synchronization at a dynamically decided frequency across the controller cluster. Hence, the maximum number of allowed concurrently executed per-replica transactions varies based on the current SDN application performance observed during runtime. The adaptation mechanism thus optimizes the trade-off between the correctness and scalability in SDN application's decision-making logic.

Until now, the AC paradigm has lacked an experimental implementation and a proof of its practicability. Furthermore, from the simulation results presented in \cite{sakicadaptive}, the overhead of the AC's state-update blocking and state-update distribution during controller operations in the congestion periods lacked a proper analysis. In this paper, we provide the insights into the realization of an AC framework that internalizes the concept presented in \cite{sakicadaptive}, and directly compare the developed framework with the SC and EC model realizations w.r.t.: i) the response time; ii) the distribution overhead; and iii) the correctness metrics. Furthermore, we present various means and design paths that can be followed to realize its adaptive component, and compare the different design options. For the comparative study, we leverage the built-in SC APIs exposed by the open-source implementation of RAFT consensus \cite{raft} in the OpenDaylight (ODL) controller \cite{odl}. We implement our AC/EC framework as an additional registry component in ODL. The framework can thus be deployed as an alternative or an addition to the existing SC framework.

We organize the paper as follows: Sec. \ref{systemmodel} elaborates the system model. In Sec. \ref{scec_models} we briefly introduce the SC and EC state synchronization models. In Sec. \ref{ac_model}, we outline the architecture of our AC framework. Sec. VI motivates the coexistence of different consistency models. Sec. \ref{ac_realization} discusses the building blocks and algorithms of the AC framework. Sec. \ref{evalmetho} outlines the exemplary load balancer implemented for the purpose of consistency evaluation. It also discusses the system setup and the evaluation methodology. Sec. \ref{eval} discusses the results of the qualitative comparison of SC, EC and AC. Sec. \ref{related_work} discusses the related work. Sec. \ref{conclusion} concludes this paper.

%\subsection{Background and problem statement}
%	- Scalability of the system hard to achieve with:
%		- Low Response Time
%		- High Availability
%	- Adaptive consistency approaches:
%		- Lack real implementations
%	- Eventually consistent approaches:
%		- Lack any guarantees or staleness limitations
% 
%\subsection{Our contribution}
%	- Realistic Framework
%	- Shaping of the adaptive consistency framework to fruition
%	- Realistic evaluation of the two approaches
%

\section{System Model}
\label{systemmodel}
\label{clustermodel}

We assume a distributed control plane model, where multiple SDN controller replicas interconnect in order to form a logical \emph{cluster}. Each replica in our system includes a default set of decision-making applications, e.g. those that make resource reservations based on routing or load balancing decisions. The replicated SDN applications and hence the controller replicas expose a set of northbound interfaces (NBIs), allowing for the acceptance of client requests. Depending on the requirements of the data synchronization, the applications hosted in different replicas are also able to process the computations given a client request in either concurrent or serialized manner. Therefore, two possibilities exist: i) processing a client request is possible only on a single replica at a time; ii) processing a client request is possible in concurrent manner in any available replica. The first method guarantees for the true serialization of resource reservations, as the decision-making is coupled with the resource state reservations. The latter method instead relies on the isolated state synchronization and convergence of resource reservation updates after the potentially concurrent decisions were made in isolated replicas.  Obviously, in the first scenario, a serialized decision-making may lead to processing bottlenecks in a highly-loaded system \cite{suh2017toward}. In contrast, we consider a more scalable method, allowing for each reservation-state-update to initiate concurrently and asynchronously at an arbitrary controller replica. It is left up to the deployed consistency model and the state-distribution mechanism to decide the actual ordering of updates.

We consider a \emph{healthy} replica an active, non-corrupt (non-buggy) controller replica that, given the latest up-to-date state, will make \emph{correct} decisions in-line with the expected SDN application design. In contrast, a \emph{failed} replica is an inactive (downed) replica that is both: a) unable to acknowledge the acceptance of state update commits distributed by remote replicas; b) unable to service new application requests. While tolerated by design, we do not evaluate partial controller failures (i.e. resulting in incorrect decision-making or faulty data store updates) but decide to focus on correctness disadvantages stemming from desychronization of controller instances. Our design does not consider Byzantine faults.

SDN controller replicas exchange their applied state-updates for the purpose of achieving high availability of the control plane. %To achieve a fail-over without the loss of state information, each update to a replica's data store must eventually be seen by all other \emph{healthy} replicas. 
The duration of the state convergence is governed by the selection of the state synchronization protocol and the corresponding consistency model. The formation of the cluster is independent of the controllers' placement, hence spatial optimization for objectives of e.g. minimized control plane response time is orthogonal to the synchronization issue. 

To guarantee a successful synchronization of the committed state-updates across all replicas, we assume a system with enabled partial synchrony and an \emph{eventual synchronous} communication model. Thus, to make progress in SC and in AC (during blocking period) systems, a guaranteed eventual delivery of each state-update to all healthy replicas is assumed. Commiting an update in these systems requires confirmations of the majority of cluster replicas \cite{sakicresponse}. Therefore, the SC system is unable to move forward in the case of the outstanding replica confirmations. We assume a \emph{fair} and \emph{robust} control channel, where given a non-partitioned network, messages sent infinitely often are delivered infinitely often \cite{panda2017scl}.

For replica failures, the eventual propagation of updates to all nodes after a controller failure (i.e. the sender's failure) can be achieved by persisting the updates to an in-memory data store and deploying a replica recovery mechanism (e.g. a watchdog mechanism) that reinitializes the controller \cite{sakicresponse}). This assumption holds for the \emph{fail-recovery} \cite{cachin2011introduction} process abstractions, which we assume in the remainder of this work. The controller instances are allowed to fail and rejoin the controller cluster arbitrarily. Before re-enabling the recovered controller, a mechanism for synchronization of missing state updates from healthy replicas (e.g. using anti-entropy or pulling of log snapshots) \cite{panda2017scl}) must have completed successfully.

\begin{figure}
		\centering
		\includegraphics[width=0.49\textwidth]{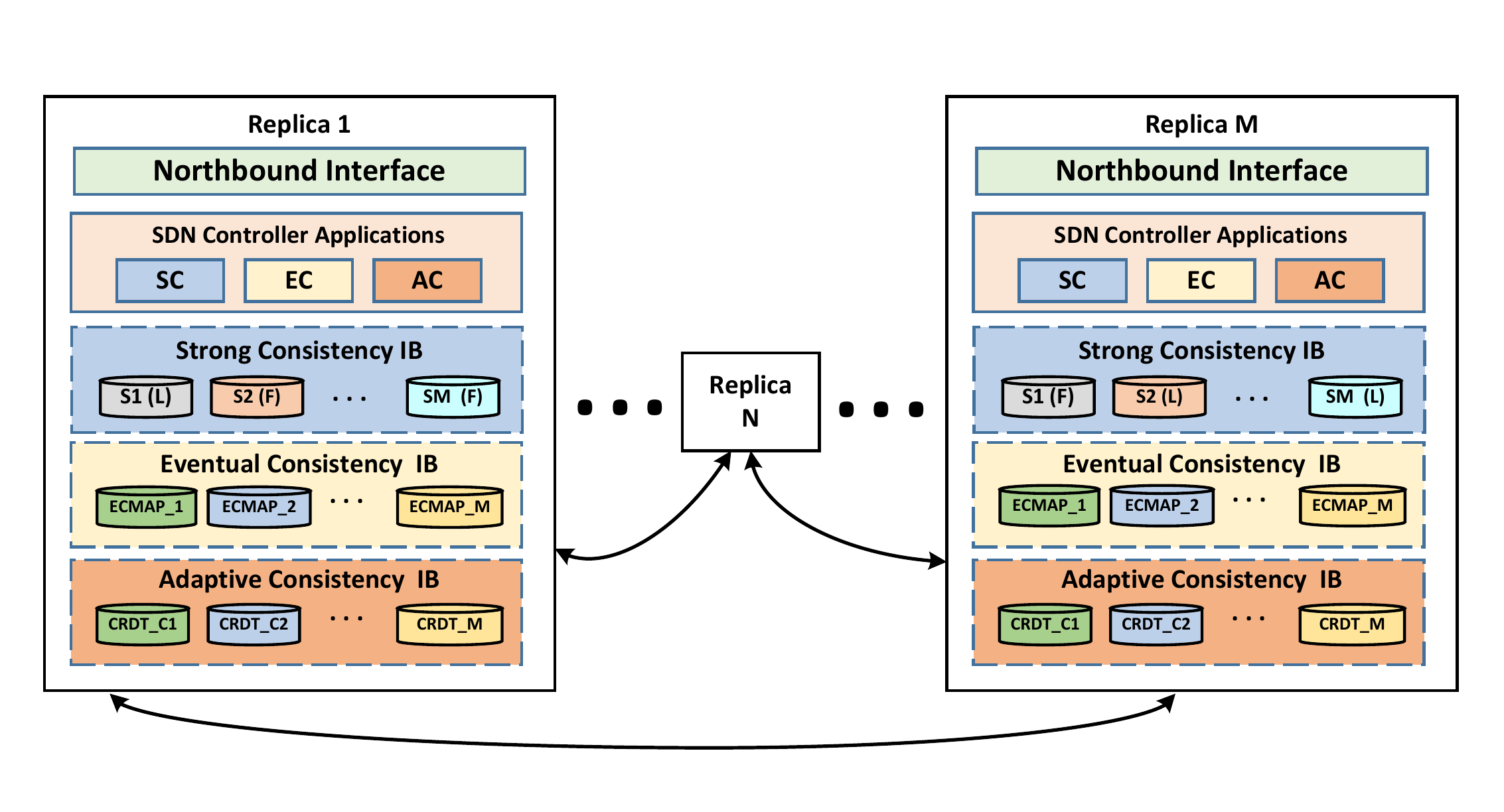}
		\caption{The internal controller model comprising data stores with varying degrees of state consistency. The controller designer should be allowed to select the appropriate data quality based on their application correctness and throughput requirements. In the case of SC, leaders are elected at a per-\emph{shard} granularity. EC and AC state instances represent exemplary \emph{EC map} realizations for the ONOS\cite{onos} EC approach and our AC approach, respectively.}
\vspace*{-0.3cm}
		\label{fig:controller_model}
 \end{figure}

Fig. \ref{fig:controller_model} presents our envisioned controller design. It depicts a number of controller replicas, interconnected for the purpose of achieving high-availability of the controller-switch and controller-client connections. Each controller executes a number of SDN applications (i.e. routing, load-balancing). Fig. \ref{fig:controller_model} depicts the case where each controller instance executes a copy of each application. For the remainder of the paper, we hold to this assumption. Thus, we allow each controller replica to execute an instance of each available SDN application (control functionality) individually. The SDN controller applications base their decisions on the current content of either one or more in-memory data store implementations which leverage different consistency models. The total controller data state in an SC cluster is partitioned into a number of \emph{data-shards}. The "sharding" of the full data store state into the data-shards is configurable at an arbitrary granularity. In OpenDaylight's realization of the SC model, an individual instance of RAFT consensus and thus the controller cluster leadership is maintained for each shard. In the remainder of the paper, for the SC model, we assume a single default data-shard replicated across each controller instance, and thus a single instance of RAFT consensus responsible for distribution of state updates. Failures of a shard leader lead to an unavailability of the read and write operations during the re-election period for the particular shard previously under the failed leader's controller. EC maps, on the other hand, are data structures whose synchronization is enforced in the background using a gossiping/broadcast primitive. In ONOS \cite{onos}, for example, EC maps are replicated to all controller instances that are members of a common cluster. Finally, the replication of the AC data structures presented henceforth is handled on a per data-state instance basis, so to allow for granular guarantees on the minimum synchronization interval of the observed data-state.

\section{Strong and Eventual Consistency Models}
\label{scec_models}

For completeness, we henceforth give a brief overview of the two common consistency models implemented in the SDN controller platforms ONOS \cite{onos} and OpenDaylight \cite{odl}.

\subsection{Strong Consistency (SC) Model}
In leader-based \emph{Strong Consistency} (SC) consensus algorithms (e.g. RAFT \cite{raft}, Paxos \cite{paxos}), each replica is assigned either a \emph{follower}, \emph{leader} or a \emph{candidate} role. Whenever a data store update is initiated by a cluster client at any active replica, the receiving controller proxies the received client request to the current cluster \emph{leader}. The leader is the controller instance that orders the incoming state-update requests, so as to allow for a serialized history of updates and thus ensure the operational state consistency during runtime. Following a committed state-update at the leader, the update is propagated to the cluster \emph{followers}. It is committed to their data store only after half of the \emph{followers} have \emph{agreed} on the update. A number of distributed consensus protocols were proposed in the past \cite{raft,ongaro2014search,pbft,paxos, viewstampedrepl, epaxos}. Currently, OpenDaylight \cite{odl} and ONOS \cite{onos} are the most attractive open-source SDN controller platforms with the largest user-/tester-base. They both provide for white-box testing and insights related to their consensus implementations. As of the time of writing this document, they both implement RAFT as the only consensus protocol. Hence, RAFT was selected as a valid implementation representative for our measurement-based study of the SC model. %In addition to a leader-based algorithm that serializes each data store update through the actual cluster master, RAFT additionally standardizes an implementation of leader election and post-failure replica recovery operations.

In contrast to the correctness benefits of the serialization of state-updates, RAFT possesses the disadvantage of an added overhead in the expected response time and lower availability \cite{sakicresponse} compared to using the eventual consistency primitives. Namely, a single data store update initiated at a follower replica requires a round trip to the leader for confirmation; as well as reaching consensus among the majority of replicas \cite{suh2017toward}.  This leads to an added blocking period and an overhead in confirmation of transactions. Furthermore, quorum-based consensus algorithms can tolerate a maximum of $F=\lceil C/2-1 \rceil$ failures in a cluster of $C$ controllers. This limitation relates to the requirement of ensuring data consistency in the case of network partitions, an invariant feasible only when a majority of nodes are involved in confirming the transactions \cite{panda2013cap, gilbert2002brewer}. In the best case, the cluster operates at the speed of the leader, and in the worst case, at the speed of the slowest follower \cite{sakicresponse}. %A comprehensive description of the algorithm can be found in \cite{ongaro2014search, Howard:2015:RRW:2723872.2723876}. 

\subsection{Eventual Consistency (EC) Model}
\label{ec_model}
\emph{Eventual consistency} (EC) claims that replicas eventually converge to the same final values independent of the applied order of operations, assuming that users (i.e. applications) eventually stop submitting new operations \cite{shapirooptimistic}. %\hl{Compared to the SC model, it provides for a superior read/write response time in most cases}\cite{levin2012logically, suh2017toward}. 
In EC, all reads and writes are performed locally at the processing speed of the local replica. Hence, applications written on top of the EC primitives proceed their operation without a penalty of confirmation time. The state-updates in EC are propagated in the background. %, thus leading to a potential data loss in the case of a controller failure during the period of state reconciliation with remote replicas. ONOS solves the issue of the potential information loss by persisting each local update to the disk, thus allowing for state reconciliation latest after the failed replica has successfully recovered. 
In ONOS, state distribution across the EC replicas occurs using an update-push distribution process and the \emph{anti-entropy}, where replicas continuously compare their local state and eventually converge the deltas. Furthermore, updates to the states may be marked with local timestamps, hence allowing for global ordering of updates. %and enabling the \emph{last-writer-wins} strategy in conflict resolution. 
EC favors the performance at the expense of consistency, potentially leading to correctness issues if the applications rely on the \emph{non-staleness} of the local state for their correct operation \cite{levin2012logically}. %\hl{possible to add routing example}

\section{Adaptive Consistency (AC) Model}
\label{ac_model}
In order to emphasize on the novelties introduced in this work, we now briefly summarize the concept and describe our realization of an \emph{Adaptive Consistency} (AC) framework. %We discuss the adaptation function, state-update distribution mechanisms, as well as the aspects of its coexistence with SC. We do so in order to provide the baseline for the evaluation of the impact of various consistency models on a distributed load balancer application in Sec. \ref{eval}.

%\subsection{General Concept}

The AC framework allows for the \emph{create}, \emph{remove}, \emph{update}, \emph{delete} operations on the granular state instances (such as counters, registers, maps etc.). The operations are \emph{eventually} synchronized between the controller replicas. However, in contrast in contrast the EC model introduced in Sec. \ref{ec_model}, that allows for enqueueing of an unbounded number of buffered unconfirmed operations, the maximum number of enqueued manipulations in an AC framework is limited by the size of an \emph{update distribution queue} and a \emph{timeout-based automated distribution} of the enqueued updates. The maximum size of the state-update distribution queue and the maximum timeout duration are governed by the currently applied \emph{consistency level} (CL). The maximum distribution queue size and the timeout are maintained at the granularity of an observed data store state. Given an application "inefficiency" metric and the optimization target, the AC framework decides on the optimal CL that is to be applied for upcoming operations.

\begin{figure}
			\centering
			\includegraphics[width=0.5\textwidth]{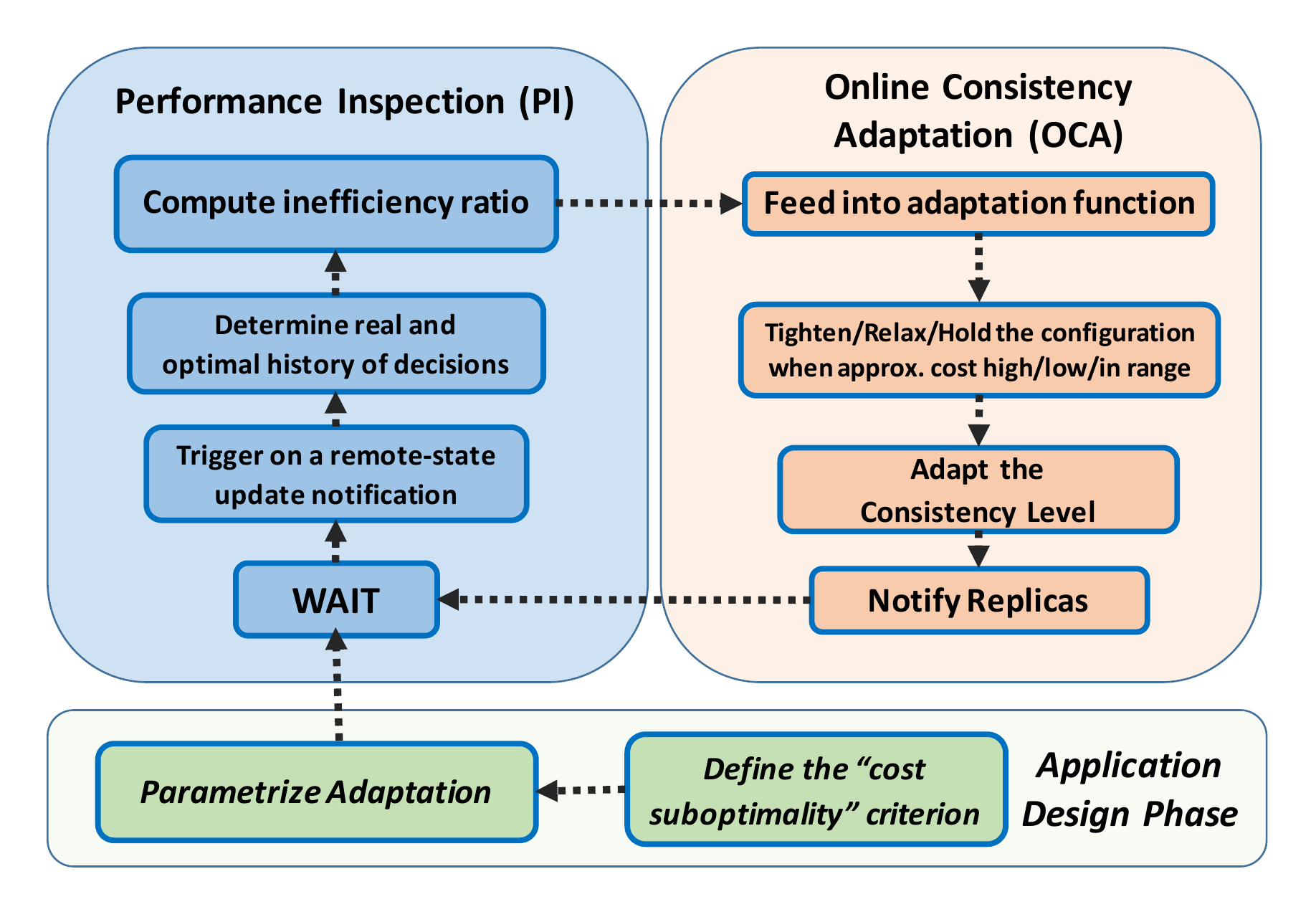}
			\caption{High-level architecture of the AC framework. The PI block is responsible for the inspection of the negative effect of the state synchronization on the quality of decision-making and the generation of a corresponding inefficiency ratio. The OCA block takes this variable as an input for the adaptation of the consistency level currently applied for the monitored state.}
			\vspace*{-0.3cm}
			\label{fig:adaptation_model}
	\end{figure}

The high-level process flow of the AC is depicted in Fig. \ref{fig:adaptation_model}. During the \emph{Application Design} phase, the designer writes an SDN application built atop of the AC framework and parametrizes the adaptation functions. During this phase, the developer must present an "inefficiency" metric related to his application logic (e.g. the optimality of routing decisions \cite{sakicadaptive}), as well as to parametrize the adaptation thresholds/efficiency targets. The specification parameters of the adaptation target vary with the choice of the adaptation function. We discuss the \emph{threshold}- and \emph{PID}-based functions in Sec. \ref{adaptations}. 

	At time $t_L$, the \emph{Performance Inspection} (PI) block triggers on an eventually delivered state-update $U$, initiated by a remote SDN controller replica $C_R$. In the PI block, each controller replica $C_L$ locally decides its own view of the real history of state-updates, by ordering the updates based on the time-stamp of updates. Let $t_U$ denote the time when update $U$ was initiated at $C_R$. Then, after deciding the global order of updates, each replica $C_L$ evaluates the effect of being \emph{late-}notified of the update $U$. To do so, it compares two histories: i) The set of results associated with the actual actions taken during the period $[t_U, t_L]$; ii) The set of results associated with the actions that would have been taken during the period $[t_U, t_L]$ \emph{if} the update $U$ had been serialized and known to $C_L$ at $t_U$. Thus, two sets of results are stored, the set of real (i.e. suboptimal) decisions, and the set of ideal (i.e. optimal) decisions. 

An \emph{inefficiency} metric (i.e. an approximation factor), estimates the ratio between the suboptimal and optimal decision, and thus the cost of eventual state-update delivery. The latest measured inefficiency is fed into the \emph{Online Consistency Adaptation} (OCA) block. In order to decide on the most-fitting CL for the observed state, the OCA block considers the latest inefficiency report as well as, optionally, the \emph{history} of previous inefficiency reports. The OCA block then decides upon the new best-fitting CL and disseminates this decision to all cluster replicas. In our AC realization, the overhead of the OCA block is centralized at a single controller that collects the remote replica's inefficiency metrics and decides on the most-fitting CL. PI and OCA blocks are pipelined for a particular state, but are parallelized for updates on different state instances, thus enabling scalable consistency adaptation.

\section{Realization of an AC Framework}
\label{ac_realization}

In this section, we present the mechanisms behind our prototype realization of the AC framework, comprising: i) a CRDT-based in-memory data store; ii) a generalized load balancer SDN controller application (SDN-LB); iii) the corresponding PI block; iv) the OCA block comprising the \emph{threshold}- and \emph{PID}-based adaptation functions; v) the mechanism for cluster-wide data store state-updates synchronization. %To our best knowledge, this is the first implementation of the AC model in the context of any available SDN controller.
%The corresponding AC framework components have been implemented in Java and integrated as a registry of the OpenDaylight controller.

\subsection{CRDT model for state-updates}

Convergent Replicated Data Types (CRDTs) \cite{shapiro2011comprehensive} are a novel approach to handling conflict-free distributed updates on a set of eventually consistent data structures. The useful property of the CRDTs is that the isolated views of a single CRDT at different replicas eventually converge to the same value, independent of the order of updates. Thus, CRDTs preserve the correctness invariant, even in the case of an increased network latency and packet loss. %(assuming retransmissions). 
With CRDTs, updates monotonically advance according to a partial order, subsequently converging towards the least upper bound of the most recent value. An example of a replicated counter datatype is a \emph{PN-Counter} (Increment/Decrement Counter), whose increment and decrement manipulations commute. Our take on a PN-Counter realization is presented in Alg. \ref{alg:crdt}. We have also leveraged CRDT register and set structures in our framework. However, for brevity we present here only the PN-Counter, and refer to \cite{shapiro2011comprehensive} for an overview of other data-types.

In our model, individual state-updates are synchronized across the SDN controller replicas, and are stored in a log-tree, together with their initiation time-stamp, for the purpose of later reference during the steps taken in the PI block. The accepted updates to a CRDT-modeled state are synchronized across the controller replicas, while rejections result in data store update failures and a subsequent notification to the requesting application. The admission control for new updates is based on the properties of the queue distribution (i.e. the maximum queue size), governed by the currently applied CL associated with the target CRDT (ref. Section \ref{sec:distribution}). 

Alg. \ref{alg:crdt} presents our PN-Counter realization. Upon a new client request to modify a particular data store object (realized as a counter CRDT instance), the local controller executes the admission control (Lines 3-9). If the update is accepted, the controller MERGEs the update with its local CRDT (Lines 13-17). It then enqueues the update for a cluster-wide distribution (ref. Section \ref{sec:distribution}). On receiving the update initiated by a remote controller, the local controller executes the \texttt{MERGE} function (Lines 22-28). Each CRDT additionally implements the \texttt{QUERY} function, allowing to read its current state.

\begin{algorithm}[htb]
	\caption{Distributed CRDT PN-Counter}
	\small
	\hspace*{\algorithmicindent} \textbf{Notation}: \\
	\hspace*{\algorithmicindent} $C_R$ Remote controller replica \\
	\hspace*{\algorithmicindent} $C_L$ Local controller replica \\
	\hspace*{\algorithmicindent} $B_j$ Client requesting a CRDT state-update \\
	\hspace*{\algorithmicindent} $Ctr_k$ PN-Counter targeted for update \\
	\hspace*{\algorithmicindent} $S_{Ctr}^{C_L}$ Set of PN-Counters stored in $C_L$'s AC data store\\
	\hspace*{\algorithmicindent} $U_{Ctr_k}$ Update request for state $Ctr_k$ \\
	\hspace*{\algorithmicindent} $B[B_j, Ctr_k]$ Update-log for client $B_j$ and state $Ctr_k$ \\
	\begin{algorithmic}[1]

		%\Procedure{Handle new counter client update}{}
		\BState \textbf{upon event} \emph{client-update}$<B_j, U_{Ctr_k}>$ \textbf{do}
			\If{$Ctr_k \in S_{Ctr}^{C_L}$}
			\State $success := evalAddToDistributionQueue(U_{Ctr_k})$
			\If{\emph{success == True}}
			\State $B[B_j, Ctr_k] \leftarrow B[B_j, Ctr_k] \cup U_{Ctr_k}$
			\State $merge(U_{Ctr_k})$
			\State $notify(B_j,$ \emph{update-success} $<U_{Ctr_k}>)$ 
			\Else
				\State $notify(B_j,$ \emph{update-failed} $<U_{Ctr_k}>)$ 
			\EndIf
			\Else
				\State $notify(B_j,$ \emph{update-failed} $<U_{Ctr_k}>)$ 
			\EndIf
			%\EndProcedure
		\\	
		\Function{merge($U_{Ctr_k}$)}{}
			\If{$U_{Ctr_k}.operation == DECREMENT$}
			\State $Decr[Ctr_k] \leftarrow Decr[Ctr_k] \cup U_{Ctr_k}.amount$
			\ElsIf{$U_{Ctr_k}.operation == INCREMENT$}
			\State $Incr[Ctr_k] \leftarrow Incr[Ctr_k] \cup U_{Ctr_k}.amount$
			\EndIf
			\EndFunction
		\\
		\Function{query($Ctr_k$)}{}
			\State \emph{return($\sum_j Incr[Ctr_k]_j - \sum_j Decr[Ctr_k]_j$)}
			\EndFunction
		\\
		%\Procedure{Handle remote-replica-initiated update}{}
		\BState \textbf{upon event} \emph{remote-update}$<C_R,<B_r, U_{Ctr_k}>>$ \textbf{do}
			\If{$Ctr_k \notin S_{Ctr}^{C_L}$}
			\State $notify(C_R,$ \emph{update-failed} $<U_{Ctr_k}>)$ 
			\ElsIf{$Ctr_k \in S_{Ctr}^{C_L}$}
			\State $B[B_r, Ctr_k] \leftarrow B[B_r, Ctr_k] \cup U_{Ctr_k}$
			\State $merge(U_{Ctr_k})$
			\State $notify(C_R,$ \emph{update-success} $<U_{Ctr_k}>)$ 
			\EndIf
		%\EndProcedure

	\end{algorithmic}
	\label{alg:crdt}
\end{algorithm}

\subsection{Performance Inspection (PI)}

The adaptation of the CLs of a particular state is based on a provided application \emph{inefficiency} metric. We define the inefficiency metric as the approximation ratio between the series of \emph{observed} and \emph{optimal} results of an SDN controller application's decisions. The \emph{optimal} result comprises the decisions the application would have made if each update in the system had been serialized (i.e. consensus-based). The \emph{observed} result is the one the local replica has achieved in an online manner, based on its own local state, and without consideration of the status of other replicas. For a replica to compute the optimal result, the knowledge about the content and timing of the eventually delivered updates must be available. The timing characteristics are necessary for the total ordering of the observed and eventually delivered state-updates.  

The generalized calculation of the inefficiency metric is depicted in Alg. \ref{alg:approxcomp}. In Lines 6-8 the PI block identifies the previously executed operations on the observed state, in the period before the \emph{remote} update $U_{Ctr_k^{remote}}$ initiation at the remote replica $C_R$. Thus, Line 8 yields an array of consistent entries which correspond to a part of the serialized true history of updates $S_{U_{cnst}}$. Lines 10-12 identify the set of client requests that have resulted in \emph{potentially} suboptimal decisions, made in the past by the local replica $C_L$, without the consideration of the eventually delivered remote state-update. Lines 14-16 then derive the application-specific optimal decisions, given the identified optimal history $S_{U_{cnst}}$, the serialized remote update $U_{Ctr_k}^{remote}$ and a set of client requests $R_{U_{incnst}}$, previously served in a suboptimal manner. The method \texttt{CompInefficiency()} in Line 20 takes as an argument the consistent (optimal) history of decisions $S_{U_{cnst}}$, and the actual, potentially suboptimal, history of decisions $S_{U_{incnst}}$. It then returns the inefficiency (approx. ratio) $\phi$ given the two series of decisions. 

Let $\sigma_{u}^R$ and $\sigma_{o}^R$ denote the cost of suboptimal and optimal decisions for a request $R$ in general case, respectively. Then, the binary value of $X_{subopt}^R$ denotes an \emph{inefficient} result, induced by the staleness (caused by delayed synchronization):

\begin{equation*}
	X_{subopt}^R = \begin{cases}
		1 & \mathrm{when}\ \sigma_{u}^R>\sigma_{o}^R\\
			0 & \mathrm{when}\ \sigma_{u}^R\le\sigma_{o}^R
	\end{cases}
\end{equation*}

\texttt{CompInefficiency()} and \texttt{AppLogic()} functions are application-specific implementations. In the next section we present an exemplary \texttt{CompInefficiency()} realization for a generalized online load balancer \cite{naik2016cost}. Its \texttt{AppLogic()} realization is assumed to optimally assign each incoming client request to the replicated server instances based on its \emph{current local} view of the server resource utilizations. We evaluate the algorithm in implementation in Sec. \ref{evalmetho}.

\begin{algorithm}[htb]
	\caption{Inefficiency calculation for a distributed CRDT}
	\small
	\hspace*{\algorithmicindent} \textbf{Input}: \\
	\hspace*{\algorithmicindent} $C_R$ Remote controller replica \\
	\hspace*{\algorithmicindent} $C_L$ Local controller replica \\
	\hspace*{\algorithmicindent} $U_{Ctr_k}^{remote}$ Reported remote update request for state $Ctr_k$ \\ 
	\hspace*{\algorithmicindent} $U_{Ctr_k}^{local}$ Local update request for state $Ctr_k$ \\ 
	\hspace*{\algorithmicindent} $S_{Ctr_k^U}$ Set of previously logged updates for state $Ctr_k$ \\ 
	\hspace*{\algorithmicindent} $U(T)$ Timestamp of the state-update $U$ at $C_R$ \\ 
	\hspace*{\algorithmicindent} $U(R)$ Client request that resulted in the update $U$ \\ 

	\begin{algorithmic}[1]
		\Procedure{Handle new counter update}{}

		\BState \textbf{upon event} \emph{update} $<C_R, U_{Ctr_k^{remote}}>$ \textbf{do}
		\State $S_{U_{cnst}} := \oldemptyset$
		\State $S_{U_{incnst}} := \oldemptyset$
		\\
		\ForAll{$U_{Ctr_k}^{local} \in S_{Ctr_k^U}$}
		\If{$U_{Ctr_k}^{local}(T) < U_{Ctr_k}^{remote}(T)$}
		\State $S_{U_{cnst}} \leftarrow S_{U_{cnst}} \cup U_{Ctr_k}^{local}$
		\EndIf	
		\EndFor
		\\
		% Determine the set of local requests that need recomputation
		\ForAll{$U_{Ctr_k}^{local} \in S_{Ctr_k^U}$}
		\If{$U_{Ctr_k}^{local}(T) >= U_{Ctr_k}^{remote}(T)$}
		\State $R_{U_{incnst}} \leftarrow R_{U_{incnst}} \cup U_{Ctr_k}^{local}(R)$
		\EndIf
		\EndFor
		\\
		\ForAll{$R_{U} \in R_{U_{incnst}}$}
		\State ${U_{Ctr_k}^{localOpt}} := \text{\emph{AppLogic}}(R_{U}, S_{U_{cnst}})$
		\State $S_{U_{cnst}} \leftarrow S_{U_{cnst}} \cup U_{Ctr_k^{localOpt}}$
		\EndFor
		\\
		\State $S_{U_{cnst}} \leftarrow S_{U_{cnst}} \cup U_{Ctr_k}^{remote}$
		\State $S_{U_{incnst}} \leftarrow S_{Ctr_k^U} \cup U_{Ctr_k}^{remote}$
		\State $\phi =$ \emph{CompInefficiency} $(S_{U_{incnst}} ,S_{U_{cnst}})$
		\\
		\State $reportIneff(\phi)$
		\EndProcedure

	\end{algorithmic}
	\label{alg:approxcomp}
\end{algorithm}

\subsection{Computation of the inefficiency metric for a generalized online load balancer SDN application}
\label{ineffratio}

For a set of defined data store states $S$ and a state-update $U(t-n)$, timestamped at time $t-n$, let $T(t-n)$ be the matrix of observations of the states encompassing the period $[t-n\ ..\ t]$. Then, $T(t-n)$ is a matrix of $|S|$ x $n$ elements.

Let $S(i)$ be the $i$th vector of the observed state values at time $t-n+i$, so that:

\begin{equation*}
\resizebox{.5 \textwidth}{!} 
{
	$\mathrm S(i) = (s_1(i), s_2(i) .. s_m(i)): S(i) \in T(t-n)$ \\
	$\mathrm{\ s.t.\ } N_{res}(i) = |S(i)|$ \\
}
\end{equation*}

%\begin{equation*}
%	\mu_{S(t)} = \frac{ \sum_{i=1}^{N_{res}(t)} s(i)}{N_{res}(t)} 
%\end{equation*}

First, let $S_{U_{incnst}}$ contain the suboptimal (real) history of state-updates. Let $S_{U_{cnst}}$ accordingly hold the computed ideal (optimal) history of state-updates (computed as per Lines 18-19 of Alg. \ref{alg:approxcomp}). Then, for each vector (time-point) $i$ of observed state values $S_{U_{incnst}}$ and $S_{U_{cnst}}$ we can compute the costs of optimal and suboptimal decisions at time $i$, $\sigma_{o}^{R_i}$ and $\sigma_{u}^{R_i}$ respectively, using standard deviation metric: 

\begin{equation*}
\resizebox{.5 \textwidth}{!} 
	{
		$\sigma_{o}^{R_i} = \sqrt{\frac{1}{N_{res}(i)} \sum_{j=1}^{N_{res}(i)} (s_j(i) - \mu_{S_{cnst}^{R_i}})^2 } \\ \mathrm{\ where\ } s_j(i) \in S_{U_{cnst}}(i)$
	}
\end{equation*}

\begin{equation*}
\resizebox{.5 \textwidth}{!} 
	{
		$\sigma_{u}^{R_i} = \sqrt{\frac{1}{N_{res}(i)} \sum_{j=1}^{N_{res}(i)} (s_j(i) - \mu_{S_{incnst}^{R_i}})^2 } \\ \mathrm{\ where\ } s_j(i) \in S_{U_{incnst}}(i)$
	}
\end{equation*}

where

\begin{equation}
	\mu_{S^{R_i}} = \frac{ \sum_{j=1}^{N_{res}(i)} s(j)}{N_{res}(i)}
\end{equation}	

	represents the mean utilization of resources at time-offset $i$, and $R_i$ represents the client request at time-offset $i$.
	%Simple inefficiency $\Phi$ at time \emph{t}:

	%\begin{equation*}
	%\Phi_{simple} (t) = \frac{\sigma_{u}(t)}{\sigma_{o}(t)}
	%\end{equation*}

Finally, after having computed the costs of optimal- and suboptimal decisions, the average inefficiency $\Phi_T$ for the observation interval $T(t-n)$ can be computed as:
\begin{equation*}
	\Phi_T = \frac{\sum_{i=0}^{\|T\|} \sigma_{u}^{R_i}}{\sum_{i=0}^{\|T\|} \sigma_{o}^{R_i}}
\end{equation*}

\subsection{Online Consistency Adaptation (OCA)}
\label{adaptations}
The OCA block is responsible for the collection of computed inefficiency values and their online evaluation. The output of the OCA block is the adapted CL for the observed state instance. The computed inefficiency value $\phi$ is input into the OCA block and the adaptation function \texttt{reportIneff()} is called, as depicted in Fig. \ref{fig:adaptation_model} and Alg. \ref{alg:approxcomp}, respectively. 

We present two methodologies for adapting the applied CL, given a historical set of inefficiency reports:

	\subsubsection{Threshold-based CL adaptation} If the observed mean inefficiency over a window of inefficiency observations of size $W$ is below, above or in between the lower and upper thresholds, the adaptation function decides to raise, lower or keep the currently applied CL, respectively. Threshold-based CL adaptation is specified in Alg. \ref{threshold-based}.
	\subsubsection{PID-based CL adaptation} In addition to the \emph{integral} part, the PID-based feedback compensator also considers \emph{proportional} and \emph{differential} parts of the recent inefficiency reports. Each part can be assigned a corresponding weight, thus allowing to favor either a fast adaptation response or long-term adaptation accuracy. For the PID-based adaptation we additionally configure the single target value the function aims to achieve at runtime. Alg. \ref{pid-based} describes the procedure.

\begin{algorithm}
	\caption{Threshold-based Consistency Level Adaptation}
	\small
	\hspace*{\algorithmicindent} \textbf{Input}: \\
	\hspace*{\algorithmicindent} $CL_{Ctr_k}$ Currently applied CL for counter $Ctr_k$\\ 
	\hspace*{\algorithmicindent} $S_{\phi}^{Ctr_k}$ Set of previously stored inefficiency reports \\ 
	\hspace*{\algorithmicindent} $T_U$ Upper adaptation trigger for the threshold metric \\
	\hspace*{\algorithmicindent} $T_L$ Lower adaptation trigger for the threshold metric \\
	\hspace*{\algorithmicindent} $W$ Window size of considered inefficiency observations \\ %in the "historical" consistency level adaptation mode. 

	\begin{algorithmic}[1]
		\Procedure{Handle new inefficiency report}{}
		\Function{reportIneff($\phi$)}{}
		%\If{$mode == "historical"$}
		\State $S_{\phi}^{Ctr_k} \leftarrow S_{\phi}^{Ctr_k} \cup \phi$
		\State $S_{Rlv} := S_{\phi}^{Ctr_k}[|S_{\phi}^{Ctr_k}|-W:|S_{\phi}^{Ctr_k}|]$
		\State $\mu_{S_{Rlv}} := \frac{ \sum_{i=0}^{|S_{Rlv}|} S_{Rlv}(i) } {|S_{Rlv}|}$
		\\	
		\If{$\mu_{S_{Rlv}} \ge T_U$}
			\State $\emph{raiseCL} \ (CL_{Ctr_k})$   
		\ElsIf{$\mu_{S_{Rlv}} \le T_L$}
			\State $\emph{lowerCL} \ (CL_{Ctr_k})$   
		\EndIf
		\EndFunction
		\EndProcedure
	\end{algorithmic}
	\label{threshold-based}
\end{algorithm}

\begin{algorithm}[htb]
	\caption{PID-based Consistency Level Adaptation}

	\hspace*{\algorithmicindent} \textbf{Input}: \\
	\hspace*{\algorithmicindent} $CL_{Ctr_k}$ Currently applied CL for counter $Ctr_k$\\ 
	\hspace*{\algorithmicindent} $S_{\phi}^{Ctr_k}$ Set of previously stored inefficiency reports \\ 
	\hspace*{\algorithmicindent} $T_O$ Target oscillation value \\
	\hspace*{\algorithmicindent} $I_g, P_g, D_g$ Integral, proportional and differential gains \\
	\hspace*{\algorithmicindent} $W$ Window size of considered inefficiency observations \\ %in the "historical" consistency level adaptation mode. 

	\begin{algorithmic}[1]
		\Procedure{Handle new inefficiency report}{}
		\Function{reportIneff($\phi$)}{}
		%\If{$mode == "historical"$}
		\State $S_{\phi}^{Ctr_k} \leftarrow S_{\phi}^{Ctr_k} \cup \phi$

		\State $P_{term} := P_g * (T_O  - S_{\phi}^{Ctr_k}[|S_{\phi}^{Ctr_k}|])$
		\State $I_{term} := I_g * \sum_{i=|S_{\phi}^{Ctr_k}|-W}^{|S_{\phi}^{Ctr_k}|} ({S_{\phi}^{Ctr_k}}_i - T_O)$
		\State $D_{term} := D_g * [(S_{\phi}^{Ctr_k} [|S_{\Phi}^{Ctr_k}|] - T_O) - (S_{\phi}^{Ctr_k} [|S_{\phi}^{Ctr_k}| - 1] - T_O)]$
		\\
		\State $T := P_{term} + I_{term} + D_{term}$
		\If{$T > T_O$}
			\State $\emph{raiseCL} \ (CL_{Ctr_k})$   
		\ElsIf{$T < T_O$}
			\State $\emph{lowerCL} \ (CL_{Ctr_k})$   
		\EndIf
		\EndFunction
		\EndProcedure
	\end{algorithmic}
	\label{pid-based}
\end{algorithm}

\subsection{State synchronization strategies}
\label{sec:distribution}

To restrict the staleness, i.e. to limit the amount of unseen updates for a particular state on diverged controller replicas, AC ensures that a reliable distribution of a bounded set of updates has occurred before a new data store transaction for the target state is allowed in the system. Each time a client requests a new state-update, we evaluate the number of previously submitted unconfirmed state-updates on the local replica. If this number is above the maximum queue size governed by the currently applied CL, the transaction is rejected. Otherwise, the state-update is enqueued in a per-state FIFO queue. Depending on the distribution strategy, we distinguish two abstractions of update-state distribution. These abstractions have their trade-offs in terms of response time and the generated update distribution load in the control plane, but they both ensure the property of limited staleness by bounding the maximum number of enqueued isolated updates per replica:%. We distinguish:}
		\subsubsection{Fast-Mode State Distribution} The first procedure of Alg. \ref{distribution-algs} realizes this distribution abstraction. If the actual occupancy of the state distribution queue is below the CL-governed threshold, the state-update is \emph{admitted} for processing (Lines 3-6), otherwise it is dropped (Lines 7-8). If admitted, the update is prepared for the distribution to the other members of the cluster. The unconfirmed updates in the system are first enqueued in the distribution queue. Any new update is merged at the tail of the queue (Line 4). The distribution procedure then serializes all outstanding unconfirmed updates and distributes these to the remote replicas (Line 5). The sender replica then waits on the asynchronous confirmations for the individual updates. After all \emph{active} cluster members have acknowledged the state-update(s), the sender removes the \emph{acknowledged} updates from the distribution queue (refer to procedure \emph{"On Acknowledgment of distribution"}). 
		\subsubsection{Batched-Mode State Distribution} The second procedure of Alg. \ref{distribution-algs} realizes this distribution abstraction. The transmission of a series of unconfirmed updates on each new update has the advantage of the lowered response time and reliability in the case where some of the previously sent out packets are lost. Nevertheless, generation of a new frame for each new state-update may cause unnecessary load if the response time is not the optimization criterion. For such scenarios, we have realized a batching queue that collects a number of state-updates (Line 4), up to the maximum amount defined by the applied CL for the particular state, and distributes these in a batch to the peer replicas (Line 7). For infrequently updated state-instances, we introduce an asynchronous timer that triggers the state-update distribution whenever a non-empty queue is not distributed for the duration of time specified by the applied CL (Lines 14-17).

\begin{algorithm}
	\caption{Fast and batched distribution of state-updates}
	\small	
	\hspace*{\algorithmicindent} \textbf{Input}: \\
	\hspace*{\algorithmicindent} $U_{Ctr_k}^{local}$ Local update request for state $Ctr_k$ \\ 
	\hspace*{\algorithmicindent} $CL_{QS}^{Ctr_k}$ Max. distribution queue size for the applied CL \\ 
	\hspace*{\algorithmicindent} $CL_{TO}^{Ctr_k}$ Distribution timeout for the applied CL \\
	\hspace*{\algorithmicindent} $Q_{Ctr_k}^E$ Distribution queue for the unacknowledged state-updates committed at the local replica \\
	\hspace*{\algorithmicindent} $QC_{Ctr_k}$ List of local state-updates acknowledged by all remote replicas \\ 
	\hspace*{\algorithmicindent} $C$ The set of remote controller replicas
	\\
	\begin{algorithmic}[1]
		\Procedure{Fast-mode distribution}{}
		\Function{evalAddToDistributionQueue($U_{Ctr_k}^{local}$)}{}
		\If {$occupied(Q_{Ctr_k}^E) < CL_{QS}^{Ctr_k}$}
		\State $enqueue(Q_{Ctr_k}^E, U_{Ctr_k}^{local})$
		\State $distribute(C, Q_{Ctr_k}^E)$
		\State \emph{return} True
		\ElsIf {$occupied(Q_{Ctr_k}^E) \ge CL_{QS}^{Ctr_k}$}
		\State \emph{return} False
		\EndIf  
		\EndFunction
		\\
		%\BState \textbf{upon event} $acknowledged<QC_{Ctr_k}>$ \textbf{do}
		%\ForAll{$U_{Ctr_k}^{local} \in QC_{Ctr_k}$}
		%\State $Q_{Ctr_k}^E\mathrm{.remove}(U_{Ctr_k}^{local})$
		%\EndFor
		\EndProcedure
	\end{algorithmic}
	\begin{algorithmic}[1]
		\Procedure{Batched-mode distribution}{}
		\Function{evalAddToDistributionQueue($U_{Ctr_k}^{local}$)}{}
		\If {$occupied(Q_{Ctr_k}^E) < CL_{QS}^{Ctr_k}$}
			\State $enqueue(Q_{Ctr_k}^E, U_{Ctr_k}^{local})$
			\If {$occupied(Q_{Ctr_k}^E) == CL_{QS}^{Ctr_k}$}	
				\State $T_{Ctr_k} == null$
				\State $distribute(C, Q_{Ctr_k}^E)$
			\EndIf
			
			\If {$T_{Ctr_k} == null$}
			     \State $T_{Ctr_k} := \emph{init-timer}(CL_{TO}^{Ctr_k})$
			\EndIf
			\State \emph{return} True
		
		\ElsIf {$occupied(Q_{Ctr_k}^E) \ge CL_{QS}^{Ctr_k}$}
			\State \emph{return} False 
		\EndIf 
		\EndFunction
		\\
		\BState \textbf{upon event} $expired<T_{Ctr_k}>$ \textbf{do}
		\State $T_{Ctr_k} == null$
		\If{$occupied(Q_{Ctr_k}^E) > 0$}
			\State $distribute(C, Q_{Ctr_k}^E)$
		\EndIf
		\EndProcedure
	\end{algorithmic}
	\begin{algorithmic}[1]
		\Procedure{On acknowledgment of distribution (fast- and batched-mode)}{}
		\BState \textbf{upon event} $acknowledged<QC_{Ctr_k}>$ \textbf{do}
		\ForAll{$U_{Ctr_k}^{local} \in QC_{Ctr_k}$}
		\State $Q_{Ctr_k}^E\mathrm{.remove}(U_{Ctr_k}^{local})$
		\EndFor
		\EndProcedure
	\end{algorithmic}
\label{distribution-algs}
\end{algorithm}

\section{Coexistence of the Consistency Models}

%\cite{panda2017scl}

%\subsection{Bounding conflict occurence: Escrow reservations}
	%- Description + algorithmic representation 

	%\subsection{Application Design considerations}
	% Configuration of the consistency levels
	% a) Manual specification
	% b) Linear transformation

	% Configuration of the threshold ratios for scaling the CLs 
	%- a) Computation of the threshold ratios (approximation ratio)	
	%	- Generalize load-balancer example

	% Scaling by conflict detection (Strong conflicts)
	%- a) Excluding conflicts (CRDTs)

	% Specification of Escrow specific data points

%}

%\subsubsection{Benefits of coupling the Strong and Adaptive Consistency Models}

A number of use cases speaks for coupling the SC and AC in a single system, specifically in the case where SC invariants may not be invalidated for only a \emph{subset} of the deployed SDN controller operations. On the other hand, AC may benefit from SC when consensus is useful for a particular non-frequent AC procedure. We henceforth name some use cases:

	\subsubsection{Policy handling with consistency invariants} Sometimes, the properties of the AC model alone are insufficient because of the strict invariant requirements. For instance, handling a routing or security policy in a consistent manner may be required when a possibility of temporary incorrect configuration exists (e.g. black holes and forwarding loops \cite{khurshid2012veriflow, zhou2015enforcing}). %Similarly, access-control configuration lists may require an atomic commit instead of the EC/AC semantics, so to ensure a correct behavior of the AAA functions. 
	Similarly, when optimal decision-making based on the current data store state is a requirement, a globally up-to-date view in each replica must be ensured at all times. 
	\subsubsection{Exactly-once semantics} Ensuring the exactly-once semantics, when multiple replicas are notified of an external event, requires consensus in order to elect the executing controller instance \cite{katta2015ravana}. Multiple triggers may lead to such events, e.g. switch state-change notifications. AC could process such events in the RSM-manner (as a Replicated State Machine \cite{schneider1990implementing}) and subsequently re-configure the switches, so that the result of the computations in the RSM instances are both compared and applied \emph{in} the switch. This, however, induces complexity that requires extended switch functionalities \cite{panda2017scl, katta2015ravana, schiff2016band}. Optionally, with a hybrid SC/AC deployment, SC mechanisms \cite{raft} could provide for the leadership semantics for the exactly-once processing on the leader, while AC would handle the subsequent state-update distribution. %Compared to the standard AC case, and similar to the SC case, this leads to a disadvantage of experiencing a temporary unavailability in the case of a leader failure \cite{sakicresponse}.
	\subsubsection{AC/SC-based CL notification distribution} Following a CL adaptation in the AC, an \emph{agreement} between the replicas is necessary to ensure the consistent global re-configuration of the CL in each controller. The agreement in \emph{all} nodes can be ensured by reaching a consensus about the newly applied CL. %or a variation of a 2- or 3-Phase-Commit protocol \cite{skeen1981nonblocking}. 
	Noted, the OCA block may execute at a single node at any point in time (e.g. the actual cluster leader) or each replica in distributed manner. The latter variant requires the nodes reaching a consensus on the new applied CL after collecting the remote replicas' respones (e.g. using a PBFT-like signalling protocol\cite{pbft}). In any case, the OCA block does not represent an SPOF in the system. %for the quorum-based consensus solutions \cite{raft, paxos} it suffices to reach agreement among the majority of replicas to make progress. These algorithms are hence unsuitable for achieving the consistent adaptation of CLs at all nodes. 

\section{Evaluation Methodology}
\label{evalmetho}

\subsection{Application model}

To present the trade-offs in deploying either Strong Consistency (SC), Adaptive Consistency (AC) or Eventual Consistency (EC) in a multi-controller testbed, we have implemented and evaluated a distributed load balancer application (SDN-LB) as a component of a modified OpenDaylight \cite{odl} distribution. The SDN-LB allows for the embedding of isolated independent services via a YANG-modeled REST interface, characterized by the type and cost (i.e. comprising a capacity requirement). Each SDN controller replica runs data store implementations for all three consistency models, and is enabled to accept new embedding requests. 

A data-plane SDN-LB has already been investigated in the past in the context of the link-load distribution scenarios \cite{levin2012logically, schiff2016band}. However, we generalize the goals of the SDN-LB to support allocation of any type of resources (i.e. bandwidth/CPU/memory) on the selected optimal service node, given a subset of the feasible candidate nodes and their current utilizations in terms of the mappable resource as the inputs. The algorithm then decides to assign the service request, under consideration of hard resource constraints, on the node deemed as optimal w.r.t. total balance of resource utilization. We adapt the algorithm defined in \cite{naik2016cost} to facilitate immediate scheduling, i.e. an online resource mapping process. %In Alg. \ref{lb}, we present our modifications to the existing algorithm. Job $Ji$ represents the cost of a particular service request (e.g. link-load or CPU/memory utilization). For the sake of complexity, we include only a single resource constraint in our implementation. In general, however, a number of constraints may be combined during the admission step (Lines 3-6) and the algorithm may be extended with a possibly weighted cost function for the optimal selection of the serving node (Lines 11-15).

We model the state of the current reservations and the available resources as in-memory state instances in our data store realizations. SDN-LB decisions are made based on the current value of these states. Upon each successful embedding, the current node utilization is updated to include the cost of the latest request. %(Line 18). 
The controller is then in charge of disseminating the local reservation update using the update-distribution and commit mechanism implemented by the underlying data store.

In the case of the SC model, every single update in the data store, to each service node, is serialized by the RAFT leader, using the consensus abstraction. In the EC and AC framework, each new resource reservation necessitates an increment or decrement update to the respective CRDT PN-Counter object (ref. Alg. \ref{alg:crdt}). Combined with the commutative increment/decrement operations, the PN-Counter ensures eventual convergence for both EC and AC models and thus represents a good data structure fit for resource tracking realization. In AC, the state-updates to the PN-Counter are queued and distributed across the cluster based on the CL-timer and maximum queue-distribution thresholds, governed by the currently applied CL. Indeed, the adaptation function (ref. Sec. \ref{adaptations}) adapts the CL, and thus manipulates the worst-case time required to synchronize the value of the counters. Thus, the adaptation affects the quality of the embedding decisions made by the replicas running the AC framework. The inefficiency metric provided as an input for the consistency adaptation $\phi$ maps to the approximation ratio introduced in Sec. \ref{ineffratio}. Finally, in the case of EC, state-updates are queued in the state-specific FIFO distribution queue and are distributed as fast as possible (i.e. excluding any waiting period). %The update distribution queues are assumed infinitely large and are thus limited only by the allocatable working memory.

\subsection{Data Store Realizations}

To empirically compare the effects of deploying the different consistency models, we have implemented and integrated in OpenDaylight the three data store variations:
	\subsubsection{Strong Consistency} The evaluated SC data store is realized using the unmodified RAFT \cite{raft} implementation atop of Java and Akka.io\footnote{Akka Clustering and Remoting - \url{https://akka.io/}} concurrency framework included in the OpenDaylight Boron-SR4 distribution. We have modeled the data-models required by the generic SDN-LB application using YANG\footnote{YANG - A Data Modeling Language for the Network Configuration Protocol (NETCONF) - \url{https://tools.ietf.org/html/rfc6020}} modeling language. We then synthesized these models into REST APIs using ODL's YANG Tools\footnote{Yang Tools - \url{https://wiki.opendaylight.org/view/YANG_Tools:Main}} compiler.
	\subsubsection{Adaptive Consistency} The implementation abstractions of the AC framework are based on the algorithms presented in Sec. \ref{ac_model}. The framework is implemented as a set of Java bundles, and has been integrated in the OpenDaylight's OSGi environment as an in-memory data store in parallel to the SC data store. In the AC environment, similar to above, we expose the data-model for the SDN-LB application using the YANG and REST APIs. Additionally, the CL adaptation as well as the distribution of the CRDT state-updates (Sec. \ref{sec:distribution}) require a new protocol definition. We have used Google Protobuf\footnote{Google Protocol Buffers - \url{https://developers.google.com/protocol-buffers/}} to describe the corresponding data structures, as well as to serialize the on-the-wire transmissions. Asynchronous replica acknowledgments are sent out to the senders in order to acknowledge the successful state-/CL-updates at the receivers. %We have furthermore implemented a variation of a \emph{Lazy Reliable Broadcast} (LRB) state distribution \cite{cachin2011introduction} to ensure a reliable state dissemination across the cluster members. The particular utility of LRB lies in the fact that even after the sender who has partially distributed its update to just a subset of replicas has failed, the updates are eventually propagated to each remaining cluster member by the correct replicas.
	\subsubsection{Eventual Consistency} The EC implementation is based on the AC framework implementation. In the AC realization, the update-distribution queue thresholds are derived from the specification of the currently applied CLs. In EC, however, the CLs hold no relevance for the state distribution, hence the maximum queue sizes of the distributed state-updates are unbounded and can thus theoretically grow infinitely for very high service request arrival rates.
%- Escrow based reservations for conflict-free upper bound updating	
%- Agreement time to see all same reservations on all controller nodes

\subsection{On topology and parameter selection}

\begin{figure}[htb]
	\centering
	\subfloat[Internet2 topology \cite{hock2014poco}]{
		\centering
		\includegraphics[width=0.23\textwidth]{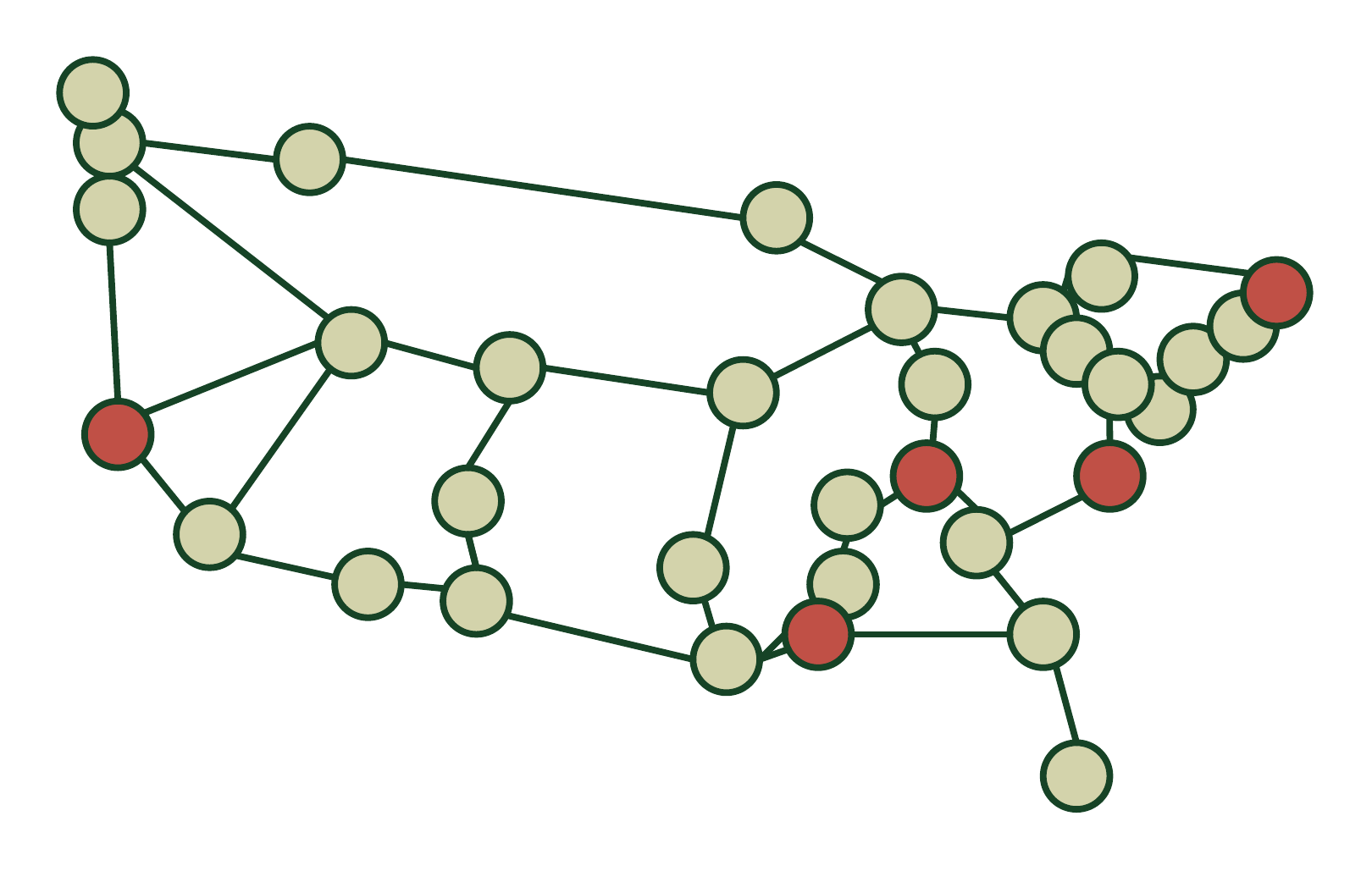}
		\label{fig:topo1}
	}
	\subfloat[Fat-tree topology \cite{huang2017dynamic}]{
		\centering
		\includegraphics[width=0.23\textwidth]{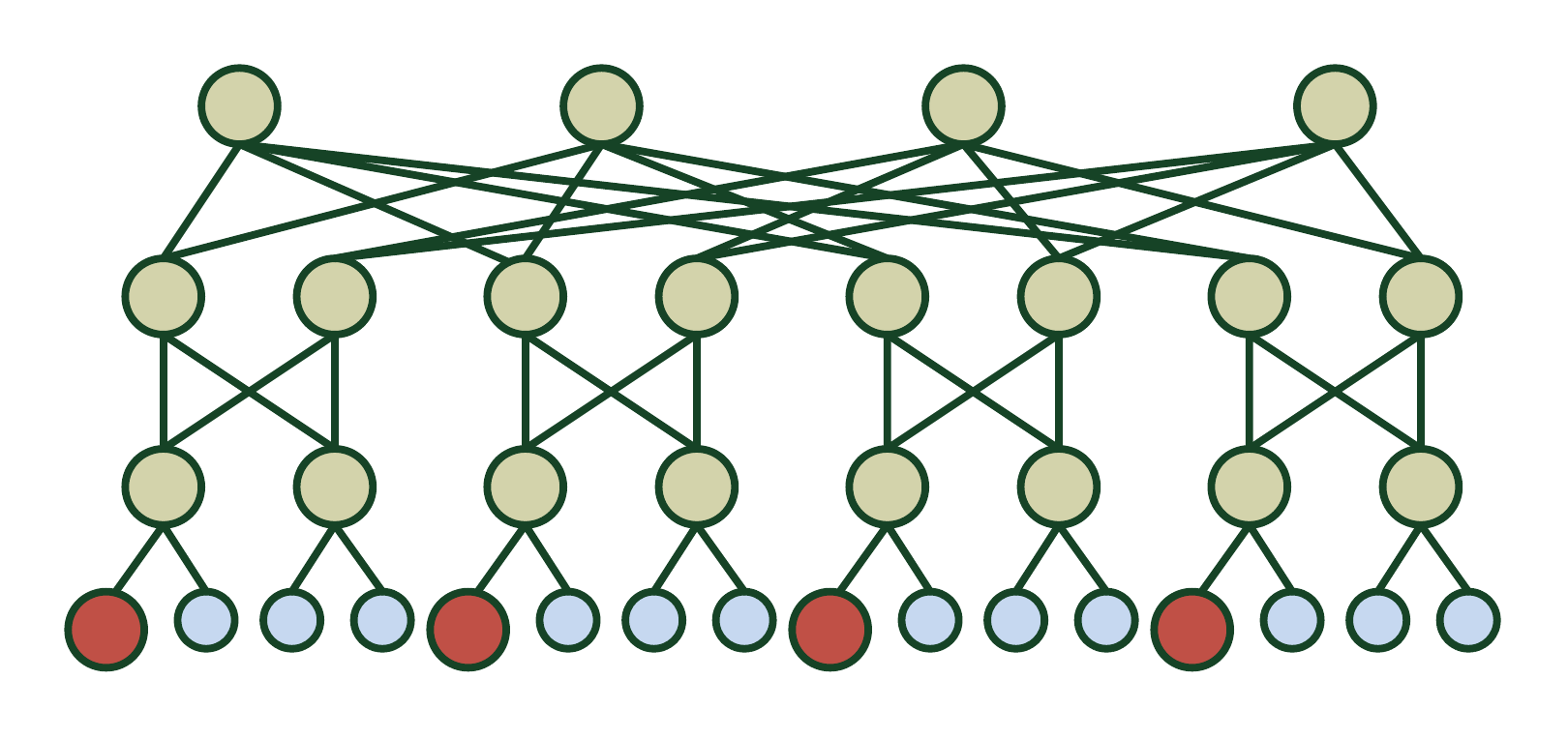}
		\label{fig:topo2}
	}
	\caption{Exemplary network topologies and controller placements used in the evaluation of the SC, AC and EC frameworks. Elements highlighted in green and blue represent the forwarding and compute devices, respectively. Red elements are the OpenDaylight controller instances placed as per \cite{hock2014poco, huang2017dynamic}.}	
	\label{topologies}
\end{figure}

We base our evaluation of the consistency models using an in-band OpenFlow control plane and an emulated forwarding plane, consisting of  a number of interconnected Open vSwitch (OVS) instances instantiated and isolated in individual Docker containers. We have emulated the Internet2 Network Infrastructure Topology as a representative of an ISP network, as well as a standard fat-tree data-center topology, controlled by a 5- and 4-controller cluster, respectively. To reflect the delays incurred by the length of the optical links in the geographically scattered Internet2 topology, we assume a travel speed of light of $2*10^6 km/s$ in the optical fiber links. We derive the link distances and hence the propagation delays from the publicly available geographical Internet2 data\footnote{Pareto Optimal Controller Placement (POCO) -\url{https://github.com/lsinfo3/poco/tree/master/topologies}}. The links of the fat-tree topology were modeled to incur a variable propagation and processing delays averaging $1ms$. In the ISP topology we leveraged a controller placement that targets the maximized robustness against the controller failures and a minimization of the probability of occurrence of switch partitions, as per the optimal placement presented in \cite{hock2014poco, lange2015heuristic}. The resulting controller placement is depicted in Fig. \ref{fig:topo1}. SDN controller replicas in the data-center topology are assumed to run on the leaf-nodes, deployed as virtual machines (VMs) (Fig. \ref{fig:topo2}), similar to the controller placement presented in \cite{huang2017dynamic}. 

%- Parameters for the Request feed-in below - data-center model as per the paper
%- Exponential inter-arrival modeling (as per paper, various frequencies)
We model the arrival rates of the incoming service embedding requests using a negative exponential distribution \cite{juan2014beyond}. To emphasize the effects of the EC on the quality of decision-making in the SDN-LB application, we distribute the total request load non-uniformly across the controller replicas. The arrival rates and the per-replica load distribution weights are included in Table \ref{parameters}. The Docker- and OVS-based topology emulator, as well as the 5 controller replicas, were deployed on a single commodity PC equipped with a recent multi-core AMD Ryzen CPU and 32 GB of RAM.

	\begin{table}
	\centering
	\resizebox{\columnwidth}{!}{%
		\begin{tabular}{ | >{\centering\arraybackslash}m{2.2cm}| >{\centering\arraybackslash}m{2cm}| >{\centering\arraybackslash}m{2cm}| >{\centering\arraybackslash}m{3.3cm}| } 
			\hline
			Parameter & Model & Value & Comment \\
			\hline 
			Number of Replicas & SC, AC, EC & [$4^{*}, 5^{+}$] & Internet2$^{+}$ and fat-tree$^{*}$ \\ 
			\hline 
			Consistency Levels (Granularity) & AC & $[1 .. 10]$ & Ref. Alg. \ref{threshold-based} and \ref{pid-based} \\
			\hline
			$CL_{QS}^{Ctr_k}$ & AC & $[3..15]$ & Ref. Alg. \ref{distribution-algs} \\
			\hline
			$CL_{TO}^{Ctr_k}$ & AC & $[100..1000]$ & Ref. Alg \ref{distribution-algs} \\
			\hline
			Initially applied CL & AC & 3 & N/A \\
			\hline
			$P_{g}$ & AC & 0.2 & Ref. Alg. \ref{pid-based}  \\
			\hline
			$I_{g}$ & AC & 0.2 & Ref. Alg. \ref{pid-based}  \\
			\hline
			$D_{g}$ & AC & 0.1 & Ref. Alg. \ref{pid-based}  \\
			\hline
			$T_{O}$ & AC &  2 & Ref. Alg. \ref{pid-based}  \\
			\hline
			$W$ & AC & 5 & Ref. Alg. \ref{threshold-based} \& \ref{pid-based} \\
			\hline
			$T_L$ & AC & 1.5 & Ref. Alg. \ref{threshold-based} \\
			\hline
			$T_U$ & AC & 3.5 & Ref. Alg. \ref{threshold-based} \\
			\hline
			$SDN-LB_{\#C}$ & SC, AC, EC & 2 & SDN-LB - No. service types \\
			\hline
			$SDN-LB_{\#S}$ & SC, AC, EC & 2 & SDN-LB - No. servers \\
			%\hline
			%$SDN-LB_{\#Samples}$ & [1000, 3000] & \\
			\hline
			$SDN-LB_{C_{Cost}}$ & SC, AC, EC & $[500..600]$ & SDN-LB - Service cost \\
			\hline
			$C_{Weights}$ & SC, AC, EC & $[1,1,2,1,5]^{+}$ $ [1,2,2,5]^{*}$ & Req. load for Internet2$^{+}$ and fat-tree$^{*}$ topologies \\
			\hline
		\end{tabular}
	}
		\caption{Parametrizations used during our study. }
		\label{parameters}
\end{table}

\section{Results}
\label{eval}
%\subsection{System resilience}
%\subsubsection{Failure Injection Model}
%\subsubsection{Time to establish control}

\subsection{Correctness of the SDN Application's Decision-Making} % compare also by assignment strategy

\begin{figure*}
	\subfloat[PID-based CL adaptation]{
		\centering
		\includegraphics[width=0.5\textwidth]{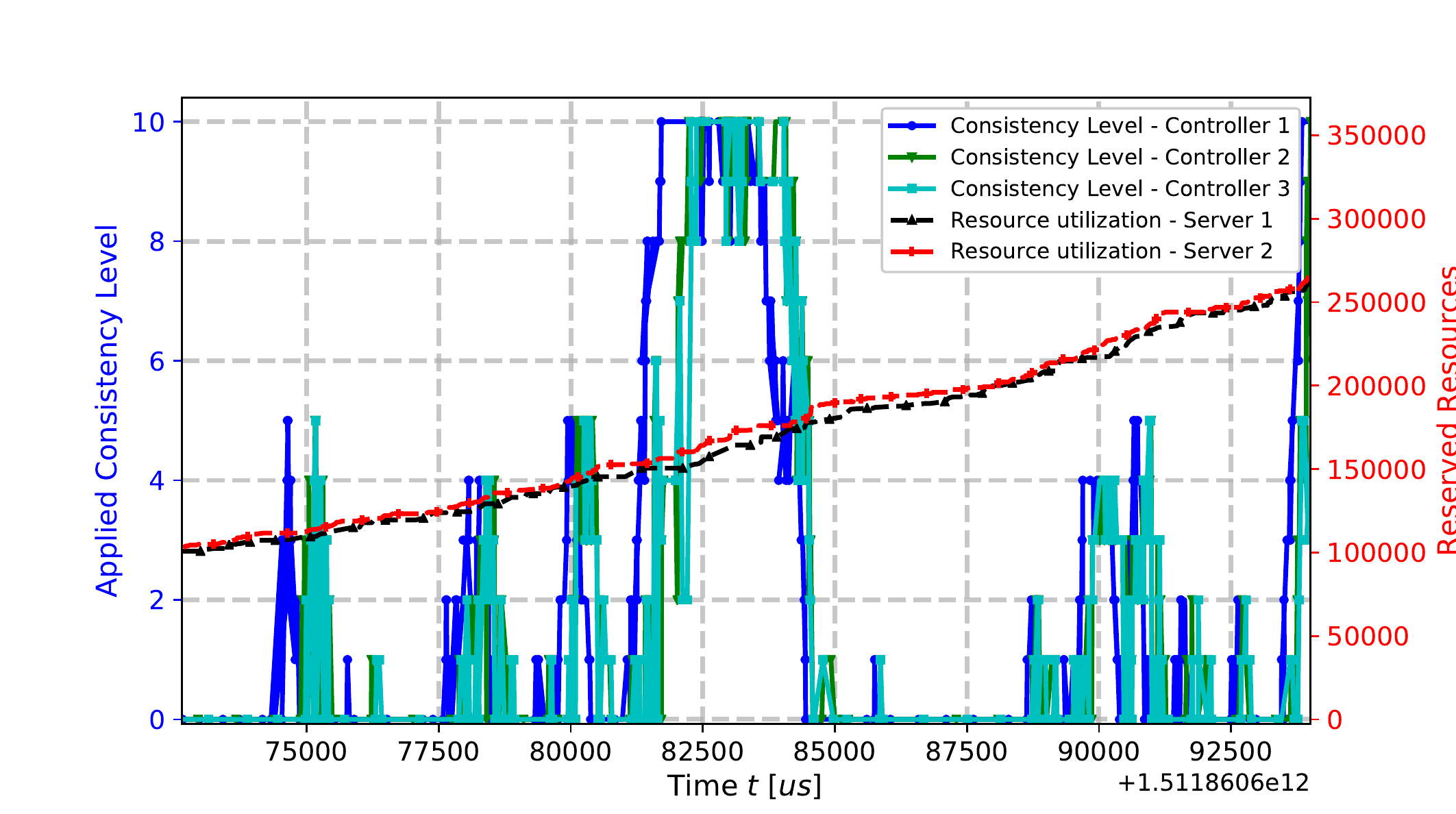}
		\label{fig:pid_adaptation}
	\label{fig:sfig1}
	}
	\subfloat[Threshold-based CL-adaptation]{
		\centering
		\includegraphics[width=0.5\textwidth]{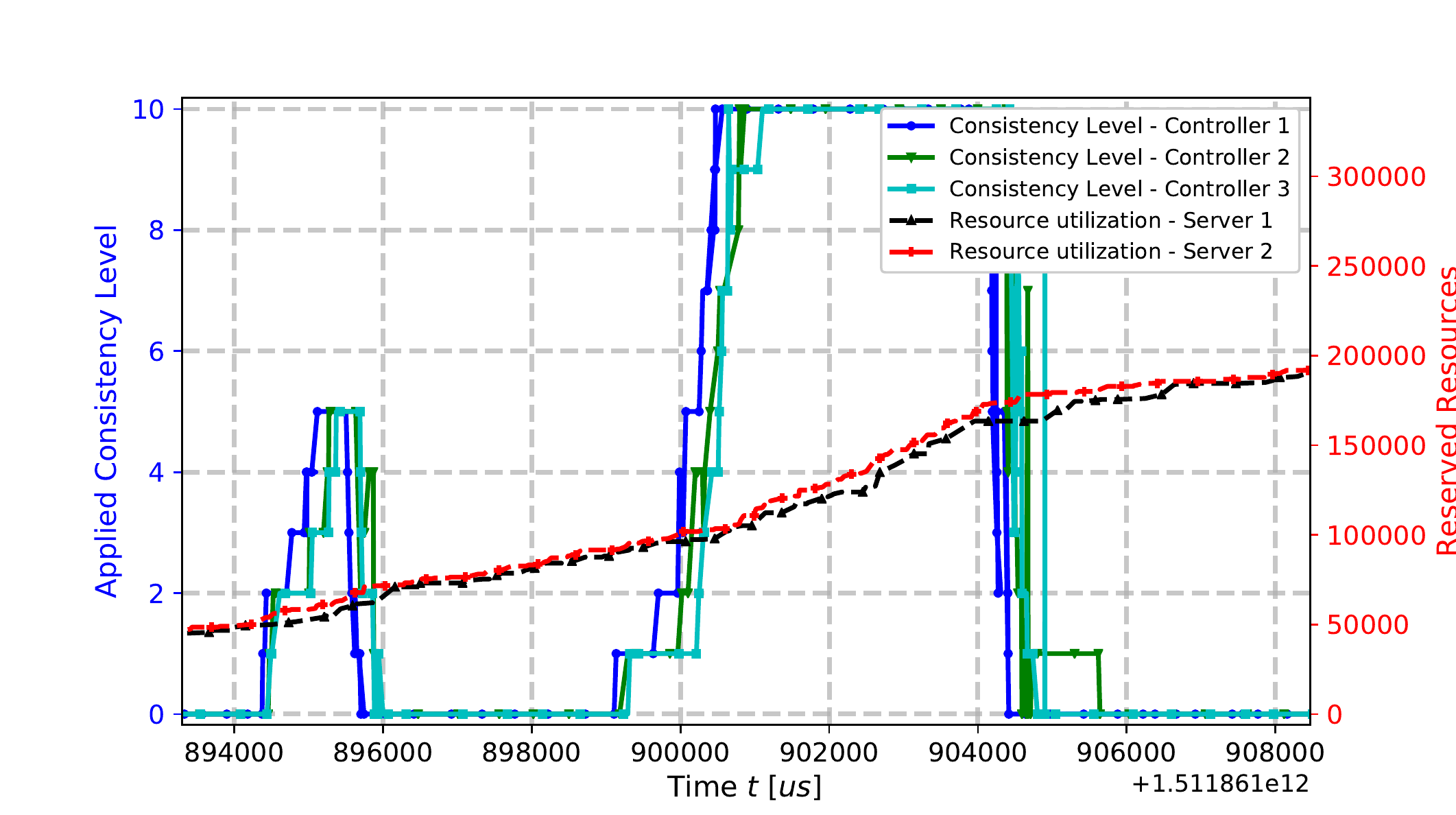}
		\label{fig:thr_adaptation}
	\label{fig:sfig2}
	}
	\caption{\emph{PID-based} (a) and \emph{threshold-based} (b) adaptation of consistency levels. The PID-based approach is more volatile compared to a rather resistant threshold-based approach. As per Alg. \ref{threshold-based} the threshold-based approach keeps the current CL unmodified, as long as the measured inefficiency stays in a specific range (i.e. between the specified upper and lower thresholds). The PID-based approach, on the other hand, oscillates around the specified target inefficiency value. For brevity, we only depict the historical data for $C=3$ controllers here (of a total of $C=5$).}
	\label{adaptation_approaches}
\end{figure*}

Fig. \ref{adaptation_approaches} visualizes an exemplary adaptation process in the AC framework. In particular, blue, green and cyan lines depict the CL applied for the SDN-LB-related CRDT PN-Counter instances on three different controller replicas. Red and black lines correspond to the actual capacity assignments managed by the SDN-LB instances on the different replicas. The resources are assigned on two different servers providing for the utilizable capacity. Indeed, in case of a strongly consistent SDN-LB (SC), the black and the red lines would continuously overlap as the state of reservations would be serialized and an optimal placement executed for each incoming service embedding request. Fig. \ref{fig:sfig2} highlights the adaptation of the CL at the time point $905000\ us$, where the imbalance and thus the inefficiency of the SDN-LB lead to an adaptation trigger and a steep decrease of the utilized CL from 10 (the most relaxed CL) to 0 (the most strict CL). The consistency adaptation function modifies the maximum available queue capacity $CL_{QS}^{Ctr_k}$ for any new state-updates as well as the worst-case timeout $CL_{TO}^{Ctr_k}$, as per Table \ref{parameters}. With the strictness of the applied CL, the SDN-LB resource assignment discrepancy decreases, but the overhead of blocking time for new state-updates increases.

\begin{figure}
	\centering
		\includegraphics[width=0.5\textwidth]{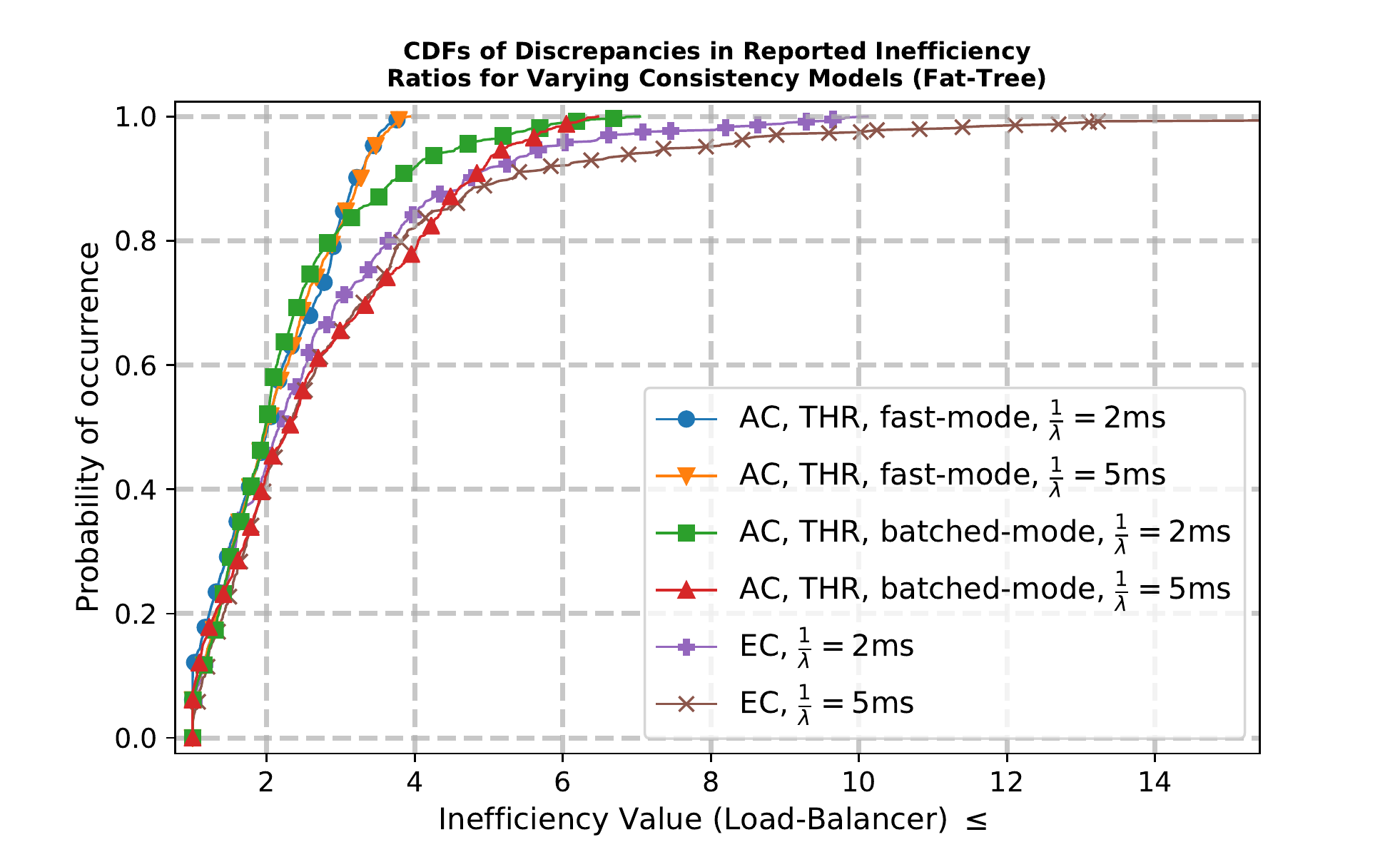}
		\caption{CDFs of reported inefficiencies (approx. ratios) for various deployed consistency models and request arrival rates in the fat-tree topology (ref. Sec. \ref{fig:topo2}). High inefficiency values indicate a more unbalanced performance of the distributed SDN-LB instances. Compared to the EC model, the AC model with the threshold-based adaptation converges faster to the worst-case with both depicted distribution modes. The fast-mode configurations result in the lowest worst-case inefficiency values. For the batched-mode distribution, the system has a similar average inefficiency as the EC. This is related to the delayed distribution of the state-updates, which is initiated only once the distribution queue is fully utilized. In the case of less frequent request arrivals (i.e. for $1/\lambda = 5ms$), the distribution queue takes the longest to fully fill up.}
	\vspace*{-0.3cm}
		\label{fig:threshold_vs_eventual}
\end{figure}

Fig. \ref{fig:threshold_vs_eventual} depicts the measured inefficiencies in the SDN-LB's assignment for the case of the AC- and the EC-based models in the fat-tree topology. The AC model uses the threshold-based adaptation of the CLs, and depicts a faster convergence in the case of the fast-mode based AC update distribution, compared to the EC case. Indeed, the fast-mode converges first to the worst-case inefficiency for all depicted request arrival rates. The batched-mode shows a lower average inefficiency than the EC model in the case of the request arrival rates of $2\ ms$. For the case of less frequent $5\ ms$ arrivals, batched-mode state-update distribution shows higher mean inefficiency than the EC case. This is explained by the fact that the batched-mode distribution of the state collects and distributes the outstanding state-updates only after a queue-threshold has exceeded or a scheduled CL-governed timer has expired. However, the time duration taken to fill up the distribution queue for the batched-mode is inversely proportional to the rate of update arrivals. Hence, compared to EC, for slow request arrivals the staleness caused by delaying the updates' distribution offsets the benefits of bounding the total number of state-updates.

\begin{figure}
	\centering
		\includegraphics[width=0.5\textwidth]{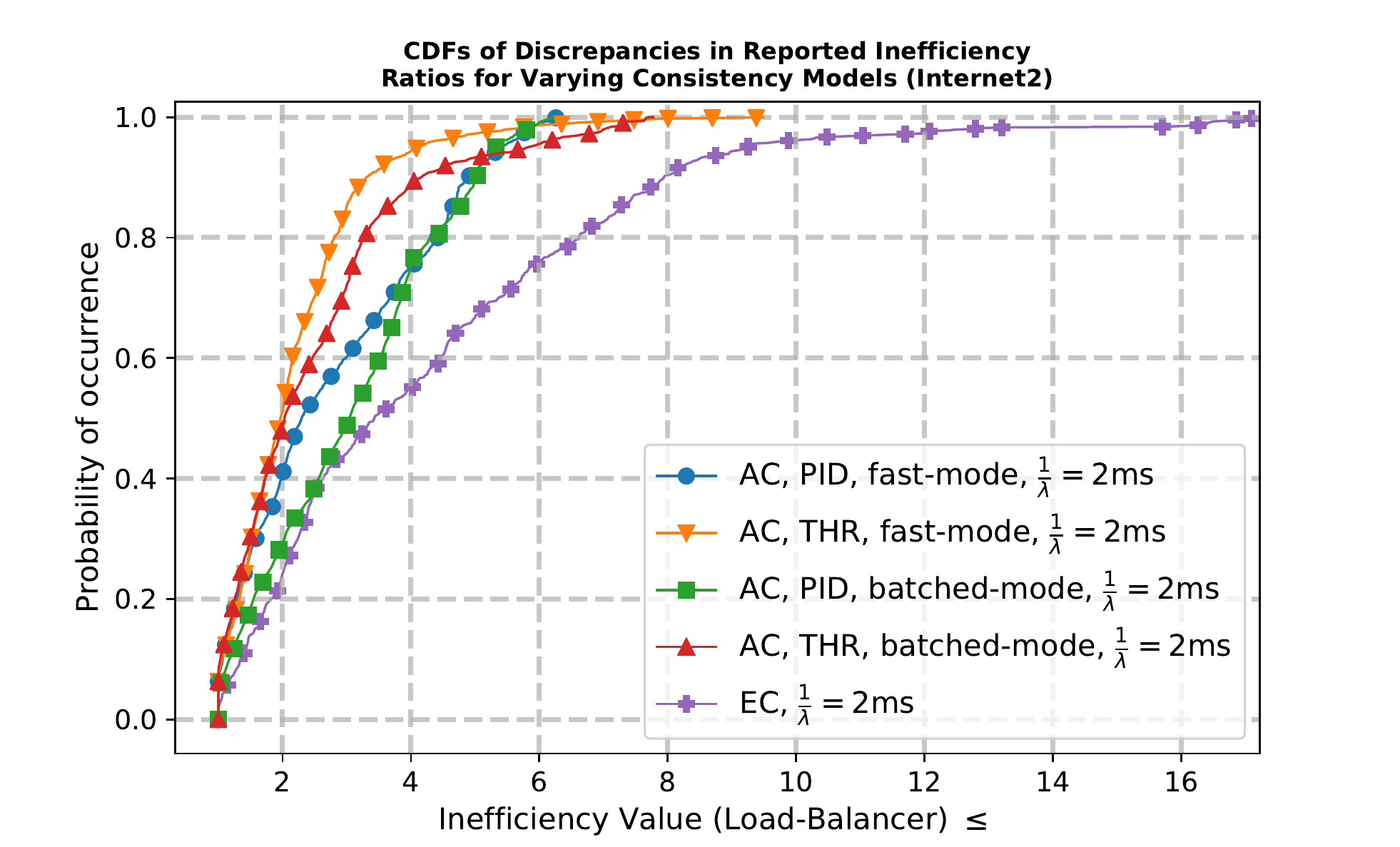}
		\caption{CDFs of the reported inefficiencies (approx. ratios) for the Internet2 topology. The average- and the worst-case inefficiencies of the EC model are larger than in the case of the fat-tree topology, as a result of the larger controller-to-controller network delays in the geographically spaced Internet2 network. The AC model deployment does not suffer from this issue because of the limited amount of incurable staleness, guaranteed by the maximum amount of unsynchronized updates. Threshold-based adaptation model shows a better performance than PID-based adaptation, possibly having to do with the higher probability of the relaxation of CLs in PID-mode (refer to Fig. \ref{adaptation_approaches}).}
	\vspace*{-0.3cm}
		\label{fig:ineff_cdfs}
\end{figure}

Fig. \ref{fig:ineff_cdfs} portrays the comparison of threshold- and PID-based AC consistency adaptation, as well as the EC consistency model in the case of the Internet2 topology. Similar to the result in Fig. \ref{fig:threshold_vs_eventual}, the usage of bounded state-update distribution queues leads to a bounded worst-case inefficiency metric with both evaluated AC adaptation functions. Interestingly, threshold-based adaptation depicts a lower average- and worst-case performance for the estimated inefficiency, compared to the PID-based model. This behavior could be caused by the tendency of the PID-model to relax the frequency of state synchronization more often, thus leading to a slightly higher inefficiency, at the benefit of a higher transaction throughput. %of state-updates and cluster-wide transactions. 

\begin{figure}
	\centering
		\includegraphics[width=0.4\textwidth]{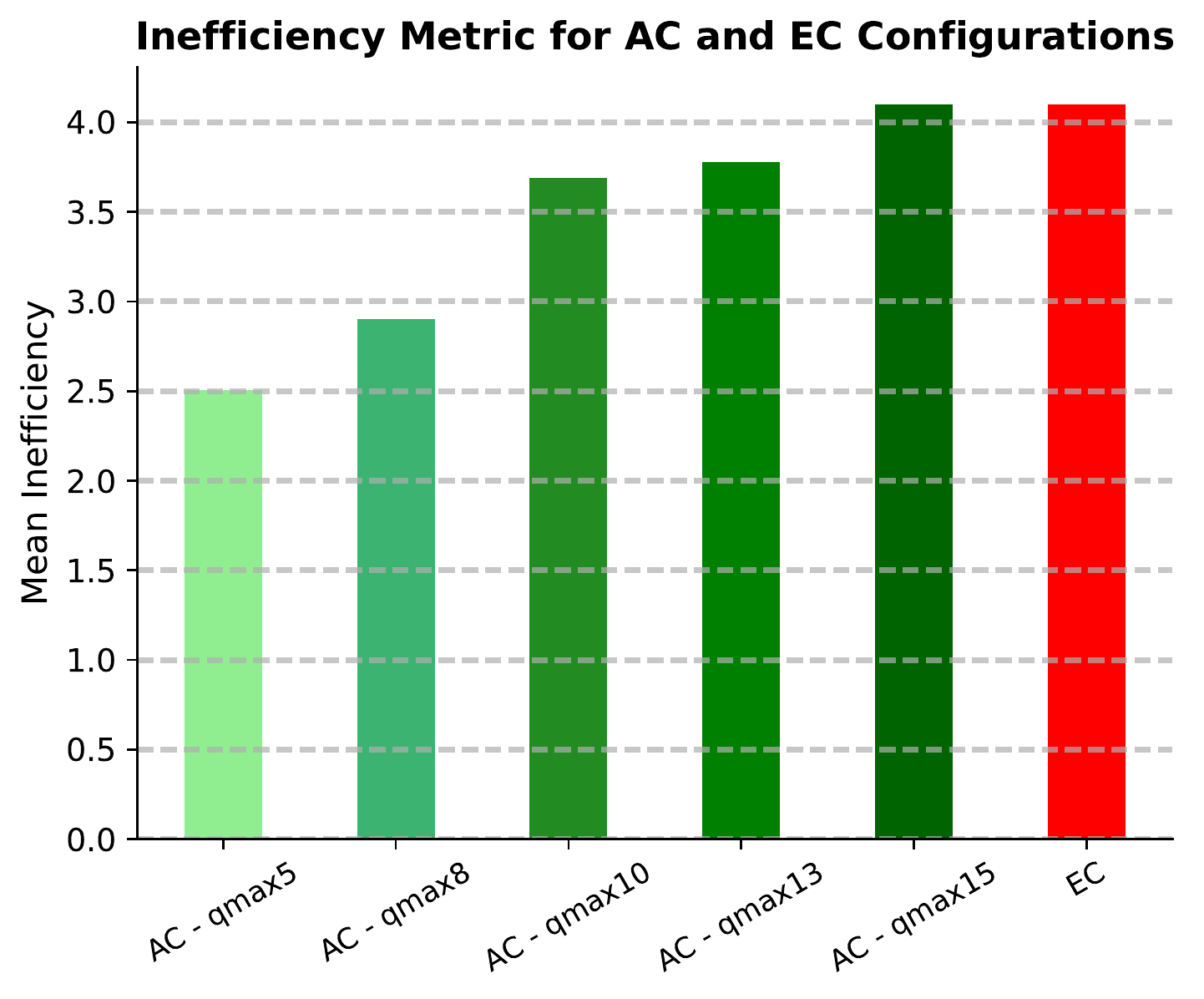}
		\caption {The measured mean inefficiency for the different $CL_{QS}^{Ctr_k}$ parametrizations for the most-relaxed CL configuration. The larger the maximum allowed state-update queue size, the larger the sojourn time of the locally admitted state-updates without enforced data store synchronization. Thus, the inefficiency of the system scales with the number of unseen state modifications. In the case of \emph{AC - qmax15}, up to 15 state-updates may be enqueued for a particular controller state without enforced synchronization. This case depicts a similar inefficiency as the EC case (without any staleness bounds). This is a result of a limited processing power of our testbed, where, at this point, clients are either unable to request a higher number of parallel concurrent REST-based SDN-LB-requests, or the controller instances are unable to concurrently process a higher number of individual updates than the 15 depicted in this case. We expect the maximum inefficiency for the EC case to deviate further in the scenarios of more scalable testbeds.} 
		\label{fig:q_effect_os3e}
\end{figure}

Fig. \ref{fig:q_effect_os3e} further emphasizes the effect of the design of CL configurations on the average-case measured inefficiency. The experienced worst-case inefficiency scales with the number of allowed isolated state-updates at a single replica. Hence, careful parametrization of the CL mappings to the maximum state-update distribution queue sizes and the timer durations is necessary to ensure the right trade-off between inefficiency and synchronization overhead.

%- Assigment strategies adaptive - batched/fast-processing
%- CDFs for various adaptation functions - pid/threshold based
%- Pid vs eventual

\subsection{The overhead of distribution of state-updates} % compare also by assignment strategy 

\begin{figure}[htb]
	\centering
		\includegraphics[width=0.4\textwidth]{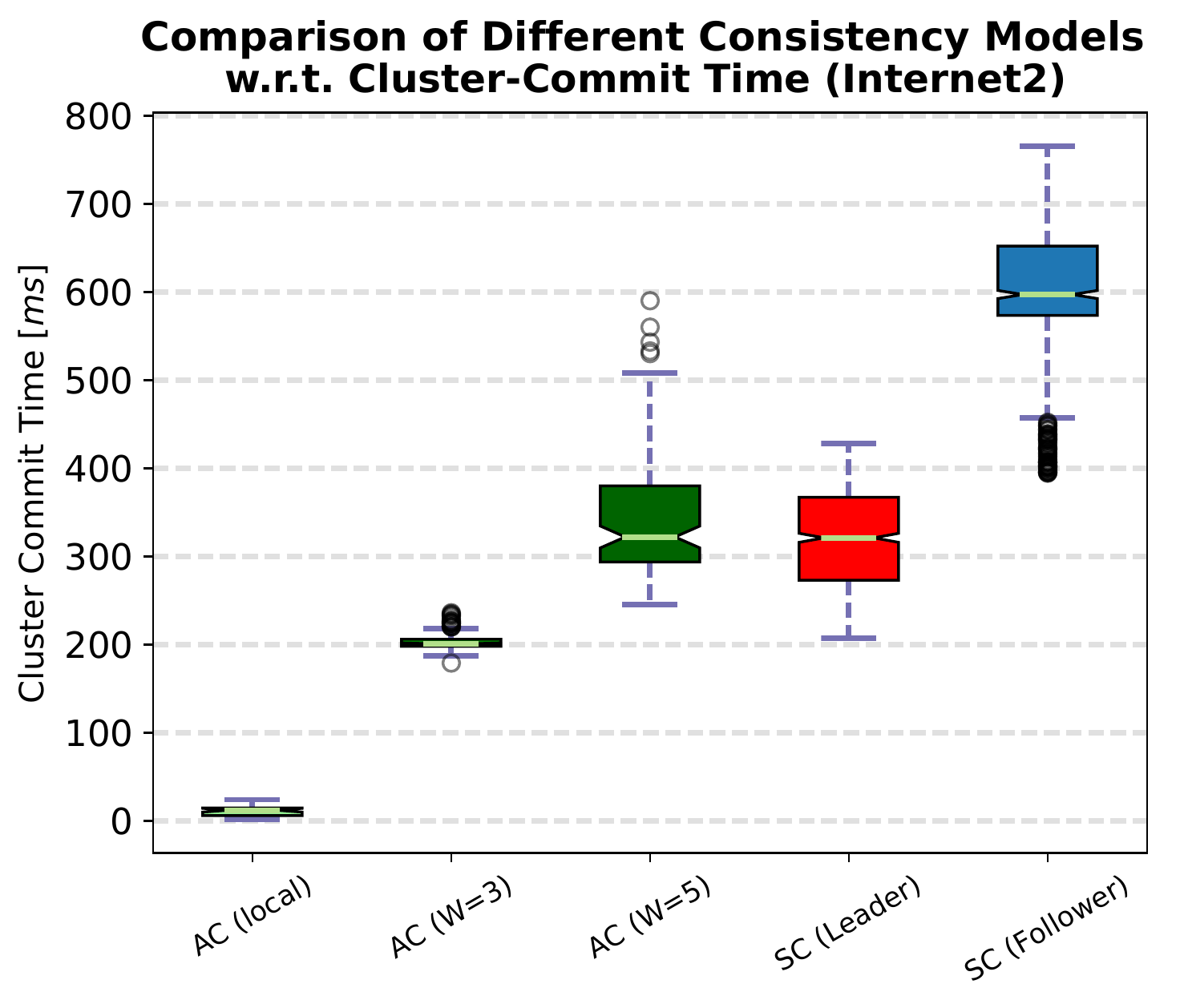}
		\caption{The commit times using the AC and SC data stores. The average response time corresponds to the \emph{AC (local)} case. The AC framework configuration that replicates to W=3 replicas \emph{AC (W=3)}, depicts a better worst-case commit-time performance, while tolerating the same availability as a comparable SC configuration (with a total of 5 controllers, and \emph{W=3} during blocking writes). The worst-cases \emph{AC (W=3)} and \emph{AC (W=5)} blocking periods, however, only occur when the state-update queue is filled and a distribution is necessary to ensure the \emph{staleness} bound. On the other hand, the throughput of the SC cluster suffers specifically in the case where updates are committed on the RAFT nodes assigned a follower role. This confirms the results of \cite{suh2017toward}.}
		\label{fig:commit_time_os3e}
\end{figure}

We define the update commit-time as the time duration required to accept, distribute and confirm a single new state-update in the underlying distributed controller data store. Fig. \ref{fig:commit_time_os3e} depicts the box-plots for the measured commit times in the case of SC and AC models. 

\emph{AC (local)} showcases the time required to apply an update to the local replica and return a corresponding acknowledgement at the requesting application. The \emph{AC (W=3)} case corresponds to the time duration needed to converge the state-update request at 3 of 5 replicas. Thus, in the case of failure of 2 controller replicas, the remaining replicas can still eventually converge on the latest state value. 

The analogue case for the SC replicas is depicted in the two right-most box-plots. An SC cluster of 5 replicas tolerates a maximum of 2 controller failures (because of the majority constraint imposed by the CAP theorem \cite{panda2013cap}). Compared to the \emph{AC (W=3)} case, the SC model offers the advantage of the serialized data stores updates. This benefit, however, comes at a high cost of minimum, average and worst case commit times, especially when state-update requests are received at one of the follower replicas. Indeed, an incoming state-update request at a leader replica leads to the faster commit confirmations, as one less uni-directional packet-transmission delay from the followers to the leader is required. The worst-case commit time in the \emph{AC (W=5)} case is similar to the optimistic SC case. It results in a relatively high commit time, because of the geographical separation of the controllers, native to the Internet2 topology. The \emph{AC (W=5)} case, however, tolerates a total of 4 controller replica failures and thus offers a higher availability compared to the SC deployment that would require a minimum of $2*4+1=9$ controller instances to tolerate the same amount of controller failures. The \emph{SC (Follower)} case considers the update requests received at one of the follower controller replicas that require transmission and subsequent serialization at leader, as well as the majority consensus to commit the update. In the geo-replicated scenarios such as in the case of Internet2 topology, this case may not be neglected.

Table \ref{overhead} portrays the incurred message load in controller-to-controller communication for the transmission of state-updates resulting from 1000 subsequent SDN-LB mapping requests, distributed uniformly across all instances. The portrayed result considers the average \emph{per-instance} overhead in a cluster of five replicas. The batched-mode in the AC framework incurs the lowest message overhead because of its useful property of aggregation of the state-updates. The SC mode depicts a lower number of frames transmitted during the per-second measurement intervals. However, the average frame size of the SC-transmitted messages is larger compared to the AC/EC deployment. The total time taken to serve 1000 SDN-LB embedding requests in the SC deployment takes a longer time as each write \emph{and} read request is serialized, and no concurrent state modifications are allowed to take place. Previous measurements of the RAFT implementation in OpenDaylight \cite{suh2016performance, suh2017toward} have proven that the overhead of read operations in a consensus-based cluster is similar to that of the write operations, since cluster-wide reads/writes are necessary to reach consensus on the latest state values. Lastly, the distribution time of AC suffers compared to the EC model, since EC processes transactions as fast as possible and does not implement the overhead of consistency adaptation.

\begin{table}
	\centering
	\resizebox{\columnwidth}{!}{%
		\begin{tabular}{ | >{\centering\arraybackslash}m{2.4cm}| >{\centering\arraybackslash}m{1cm}| >{\centering\arraybackslash}m{1cm}| >{\centering\arraybackslash}m{1.5cm}| >{\centering\arraybackslash}m{1.2cm}| } 
			\hline
			Consistency Model & Distr. Time [s] & Avg. PPS & Avg. Packet Size [B] & Avg. Load [B/s] \\
			\hline
			EC & 25.12 & 3802 & 124.5 & 469k \\ 
			\hline 
			SC & 115.12 & 326.82 & 2612.6B & 844.2k \\ 
			\hline 
			AC (Batched-Mode) & 39.501 & 3460 & 105.4 & 365k \\
			\hline
			AC (Fast-Mode) & 40 & 3552 & 102.5 & 372k \\
			\hline
		\end{tabular}
		}
		\caption{Per-replica load when serving 1000 SDN-LB requests in a 5-controller cluster synchronized using SC, AC or EC. }
		\label{overhead}
\end{table}

\section{Related work}
\label{related_work}
%SC
%In this section, we present the related work in the context of SC, AC and EC models and the existing performance evaluations.
%
\subsubsection{On Strong Consistency}

Ongaro et al. \cite{ongaro2014search} and Howard et al. \cite{raft} provide the initial experimental performance evaluations of the RAFT consensus algorithm. They focus on the evaluation of the performance of leader election procedure during the controller failure scenario. Suh et al. \cite{suh2016performance, suh2017toward}  experimentally measure the throughput and the recovery time of a RAFT-enabled cluster with up to 5 SDN controllers for the use case of flow table reconfigurations. These works do not discuss the effect of failures on the quality of decision-making in the context of SDN applications nor do they cover the aspects of RAFT scalability for high-throughput applications.

Ravana \cite{katta2015ravana} is a proposal for a distributed SDN controller that provides for a total-order of processed control messages, and ensures exactly-once delivery invariant for switch (re-)configurations. The focus of Ravana is on ensuring the performance guarantees in the face of failures, and not on leveraging the different consistency models for supporting the scalability of the distributed SDN control plane.

In our measurements, we continuously assume the availability of strict serialization and thus the exactly-once and total-order semantics in the SC model. However, a recent research \cite{zhang2017raft} has showcased two scenarios where the interplay of a RAFT-enabled controller cluster and SDN data plane may introduce inconsistency in the control plane. The authors formulate and reproduce the problems of the oscillating and non-converging RAFT leader election, and propose a partial solution. However, they leave the validation of the solution for future work. Thus, even if we compare the AC approach to an idealized SC scenario, we note that RAFT still may lead to poor correctness/availability performance in some cases. %We additionally note that a formal verification of the RAFT-correctness in the context of a distributed SDN control plane could prove a potential fruitful area of future research.

\subsubsection{On Eventual Consistency}

HyperFlow is an EC publish-and-subscribe data-broker approach \cite{hyperflow}. In HyperFlow, each controller sends the state-update requests to an external data store that disseminates the state-updates to other controllers. HyperFlow centralizes the state collection and distribution entity in the external data store, thus effectively shifting the issue of SPOF from the context of an SDN controller to another centralized instance and not resolving the SPOF itself. We focus on the internalization of the data store to stay compatible with the current clustering solutions. SCL \cite{panda2017scl} provides a methodology for preserving safety and liveness invariants without deploying consensus. The authors rely on the \emph{quiescent} period where, during a period with no data plane changes, all controllers eventually converge to same view to ensure correct operation. This can, however, only be guaranteed in networks with very limited reconfiguration dynamics, which is why SCL may occasionally lead to a disagreement of controller views. Our AC concept does not rely on quiescent period. Similarly, we do not rely on the availability of switch agents to guarantee the exactly-once execution semantics, which in contrast SCL does. %SCL also comes with a large computational overhead, as each data-plane change event leads to an \emph{isolated} re-computation of liveness and safety policies in \emph{each} controller instance.}

Levin et al. \cite{levin2012logically} evaluate the impact of an inconsistent global network view on the load balancer's performance assuming flexible frequencies of synchronization periods. Their results suggest that an inconsistency in the SDN control state view across multiple controller instances significantly degrades the performance of the SDN applications agnostic to the underlying state distribution. Contrary to our work, the authors generalize the synchronization procedure to a periodic task with flexible periods. In the case of SC, we consider a continuous synchronization model where on-the-wire transactions must be serialized in order to ensure a total ordering of decisions. In the EC case, we assume non-periodic state synchronization, as this provides for a more realistic and better performance and lower penalty of state staleness, especially for the case of higher request arrival rates. %Finally, the above authors do not consider any bounds on the total number of isolated updates in the case of applied EC consistency model, which we do using the AC semantics.

\subsubsection{On Adaptive Consistency}

In \cite{sakicadaptive}, we have introduced an AC model that employs the concept of 'strong eventual consistency', along with a 'cost-based' approach for quantifying penalties induced by the successfully detected state-merge conflicts. We used simulation to evaluate our model on an example of an SDN routing application and have motivated the potential performance gains. %when deploying resource-constrained applications. %We now extend our simulation-based study to consider and experimentally quantify the advantages of: i) the correctness constraints over the EC model; ii) the execution time and throughput benefits over the SC model; iii) the benefits in the imposed synchronization load over either EC or SC models; and iv) the benefits of the AC model on an example of an SDN-LB application. Hence, we consider this work a useful and necessary extension of our previous research.

In \cite{aslan2016adaptive}, the authors compare an adaptive approach to the state synchronization between the controllers with an approach of using the non-adaptive controllers that synchronize state with a constant synchronization period. The authors deploy an adaptation module to apply one of the pre-configured fixed synchronization intervals, which makes the approach inflexible for frequent network changes (i.e. varying controller request loads and network congestions). In contrast, we define an adaptation function which manages the new update admission in order to preserve the worst-case staleness bounds. Our work extends the conclusions on the practical applicability of AC by considering constant re-adaptation of the consistency levels.%, which is of crucial importance for non-predictable traffic models. %Furthermore, we include a realistic comparison of three different consistency models in an experimental OpenDaylight-based setup, both in the terms of the result correctness, and message and delay overhead imposed by the different models. %Finally, we present different AC realization options for the distribution and consistency adaptation procedures. 

TACT middleware \cite{yu2000design} enforces consistency bounds among the replicas of a distributed system. To bound the level of inconsistency, TACT defines consistency measures, including: i) the \emph{order error}, which limits the number of tentative writes that can be outstanding at any replica; ii) the \emph{numerical error}, which bounds the difference between the value delivered to the client and the most consistent value; iii) and \emph{staleness}, which places a probabilistic real-time bound on the delay of propagating the writes. A well-known arrival rate for the incoming requests is used to estimate the probabilistic staleness bound. %In \cite{krishnamurthy2002adaptive}, based on a requested staleness characteristic, the clients are additionally assigned a particular serving instance to handle future client requests. %Additionally, these works present a probing mechanic for more accurate estimation of replica's knowledge state staleness, at the expense of additional imposed system load. 
We distinguish ourselves from TACT by introducing an admission control mechanism for serving new state modifications at any randomly selected serving instance. Thus, we proactively place deterministic bounds on the maximum number of allowed isolated local state updates. %Furthermore, we introduce the adaptation functions for flexible and autonomous adaptation of the consistency levels given an application monitoring metric. 

A corner case of the exceeded limit values for isolated PN-Counter state-updates is related to the issue of conflicting over-reservations, previously discussed in \cite{sakicadaptive}. Balegas et al. \cite{balegas2015putting, balegas2015extending} solve this issue by implementing an escrow-based bounded counter CRDT, so to guarantee that a value of a counter never exceeds some limit value. %The authors demonstrate an explicit example, where a state value never becomes negative, even when multiple updates for the same counter are executed in different replicas.

\section{Conclusion and Outlook}
\label{conclusion}

This works presents a realization of an Adaptive Consistency (AC) framework. %We first introduce the general concept of AC and specify the different adaptation functions and state-distribution mechanisms. 
On an example of 5-controller SDN cluster and two realistic network topologies, we highlight its advantages w.r.t.: i) the control plane response time compared to the Strong Consistency approach; ii) the decision-making efficiency compared to the Eventual Consistency approach; iii) the generated controller-to-controller load compared to the both approaches. %The proposed in-memory data store leverages a middle-ground consistency for developing scalable and flexibly-correct SDN applications. %We bound the application-knowledge staleness by limiting the number of unseen remote-initiated state-update transactions. Our framework does so by means of an admission control mechanism for incoming data store updates. 
We introduce two distribution abstractions that enable the controller-to-controller state exchange, while minimizing the response time and the generated average load. Furthermore, we present two adaptation functions, that adapt the system in a closed-loop manner, so that the target inefficiency incurred by the eventuality of state delivery persists according to the SDN application's expectations.

Future works should evaluate the AC framework in scenarios comprising a larger number of controllers. Large-scale demonstrations could additionally emphasize the benefits over the alternative consistency approaches. %Furthermore, the effects of the replica failures and replica recovery on the correctness of the AC distribution mechanisms require additional investigation. %The final demonstration of the practicability of the AC concept should also consider applications beyond the constrained routing and load balancing investigated so far. 
The adaptation functions take as input the SDN application-triggered inefficiency reports. The benefit of the consistency adaptation comes at the expense of an expanded model parametrization space and the necessity of an SDN application re-design. Further attention should thus be given to simplifying the related development efforts, e.g. by providing sane configuration defaults or by automating the generation of required parametrizations \cite{Aslan2017ACC}.

\section*{Acknowledgment}

This work has received funding from the European Union’s Horizon 2020 research and innovation programme under grant agreement number 780315 SEMIOTICS and in parts under grant agreement number 647158 FlexNets (by the European Research Council). We are grateful to Majda Glotic, Dr. Johannes Riedl, Valeh Sabziyev, and the reviewers for their useful feedback and comments. 


%\bibliographystyle{IEEEtran}
\bibliographystyle{IEEEtran}
\bibliography{IEEEabrv,qos}

%\vfill


\vspace{-9mm} 
        \begin{IEEEbiography}[{\includegraphics[width=0.9in,height=1.25in,clip,keepaspectratio]{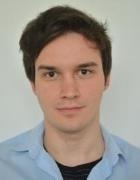}}]{Ermin Sakic} (S$'$17) received his B.Sc. and M.Sc. degrees in electrical engineering and information technology from Technical University of Munich in 2012 and 2014, respectively. He is currently with Siemens AG as a Research Scientist in the Corporate Technology research unit. Since 2016, he is pursuing the Ph.D. degree with the Department of Electrical and Computer Engineering at TUM. His research interests include reliable and scalable Software Defined Networks, distributed systems and efficient network and service management.
%He is currently involved in the EC's H2020 5G-PPP project VirtuWind. 
\end{IEEEbiography}
\vspace{-9mm} 
\begin{IEEEbiography}[{\includegraphics[width=1in,height=1.25in,clip,keepaspectratio]{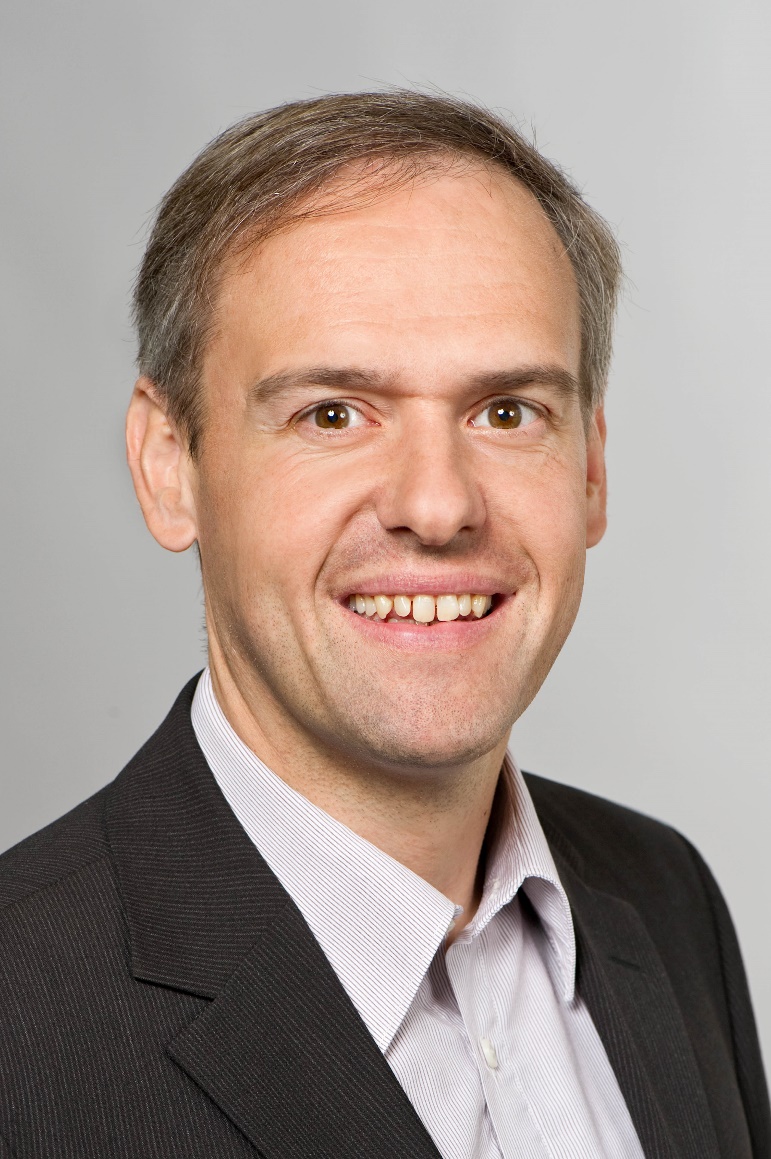}}]{Wolfgang Kellerer}
(M$'$96 \textendash\ SM$'$11) is a Full Professor with the Technical University of Munich (TUM), heading the Chair of Communication Networks at the Department of Electrical and Computer Engineering. Before, he was for over ten years with NTT DOCOMO's European Research Laboratories. He received his Dr.-Ing. degree (Ph.D.) and his Dipl.-Ing. degree (Master) from TUM, in 1995 and 2002, respectively. His research resulted in over 200 publications and 35 granted patents. He currently serves as an associate editor for IEEE Transactions on Network and Service Management and on the Editorial Board of the IEEE Communications Surveys and Tutorials. He is a member of ACM and the VDE ITG. \end{IEEEbiography}

\end{document}